\documentclass[12pt]{book}
\usepackage{a4wide}
\usepackage{scrpage}
\usepackage{german}
\usepackage{amsmath}
\usepackage{epsfig,psfrag}
\usepackage{layout}

\newcommand{\bemerkung}[1]
 {}

\renewpagestyle{headings}{(\textwidth,1pt){\headmark\hfill}{\hfill\headmark}%
  {\hfill\headmark\hfill}(\textwidth,.4pt)}%
{(\textwidth,0pt){\pagemark\hfill\today\hfill}{\hfill\today\hfill\pagemark}%
  {\hfill\pagemark\today\hfill}(\textwidth,0pt)}
\renewpagestyle{plain}{(\textwidth,0pt){\hfill}{\hfill}
  {\hfill}(\textwidth,0pt)}%
{(\textwidth,0pt){\pagemark\hfill\today\hfill}{\hfill\today\hfill\pagemark}%
  {\hfill\today\hfill\pagemark}(\textwidth,0pt)}
\def\today{}

\def\x{\vec x}
\def\tx{\vec {\tilde x}}
\def\zab{\left.z^{\alpha\beta}\right.}
\def\Sum#1{\left.\sum_{#1}\right.^>}
\def\H{{\cal H}}
\def\F{{\cal F}}
\def\Z{{\cal Z}}
\def\q{\vec q}
\def\ap{a_{||}}
\def\as{a_{\perp}}

\begin{document}

\thispagestyle{empty}
\begin{center}
  {\Huge
    The roughening transition of interfaces in disordered media}

Diploma thesis\\
\vspace{1cm}

Uwe M\"ussel\\
\textit{Institut f"ur Theoretische Physik, Universit"at zu K"oln,
    Z\"ulpicher Str. 77, 50937 K"oln, Germany}

\end{center}
\vspace{2cm}

\begin{center}
{\bf Abstract}

\begin{minipage}[h]{13cm}
Competing pinning effects on a $D$-dimensional interface by weak
impurity disorder and a periodic potential of the underlying crystal
lattice are analyzed for $2<D<4$. We use both the Gaussian variational
method (GVM) and the functional renormalization group $\epsilon=4-D$
expansion (FRG) which yield different phase diagrams: Whereas the FRG
always predicts a rough phase with irrelevant lattice pinning, the GVM
in combination with a three parameter RG for the random potential [T.
Nattermann, H.  Leschhorn, Europhys. Lett. {\bf 16} (1991) 603] leads
to a roughening transition of first order. For random bond disorder we
compute self-consistently the effective lattice potential.
\end{minipage}
\end{center}

\newpage

\title{\sffamily\bfseries\Huge
  Der Rauhigkeits"ubergang von
  Grenzfl"achen in ungeordneten Medien} 

\author{Diplomarbeit\\ von \\ Uwe M"ussel \\ \\ Institut f"ur
  Theoretische Physik \\ {Universit"at zu K"oln}}

\date{\vfill Dezember 1996}

\maketitle
\thispagestyle{empty}

\frontmatter
\tableofcontents

\mainmatter

\chapter*{Einleitung}
\addcontentsline{toc}{chapter}{Einleitung}
\markboth{Einleitung}{}

Oberfl"achen und Grenzfl"achen spielen in der gesammten
(naturwissenschaftlichen) Welt eine gro"se Rolle. Beispiele hierzu
gibt es aus der Biologie, der Chemie und, was uns besonders
interessiert, aus der Physik.
Der wichtigste Vertreter aus der Biologie ist die Zellmembran, die es
"uberhaupt erm"oglicht, das es Leben gibt.
Aus der Chemie ist sicherlich der Katalysator ein wichtiges Beispiel,
den wir alle aus dem Auto kennen. Er beschleunigt chemische Reaktionen,
indem er, durch Anlagerung der Schadstoffe an der Oberfl"ache, es
erleichtert, deren chemische Bindungen aufzubrechen.  So k"onnen neue,
weniger sch"adliche Stoffe enstehen. Sein Wirkungsgrad ist 
proportional zur effektiven Oberfl"achengr"o"se.

Aus der physikalischen Sicht ist eine Oberfl"ache ein Objekt, welches
zwei Teile eines Systems voneinander separiert. Die Dimension der
Oberfl"ache ist immer um eins kleiner als die des zugrundeliegenden
Raumes. Die Mathematiker sprechen in einem solchen Fall von einer
Hyperfl"ache. Beispiele gibt es ebenso unterschiedliche, wie
zahlreiche: So gibt es ``gew"ohnliche'' Oberfl"achen wie die
Begrenzungen eines Festk"orpers. Die Grenze zwischen zwei
koexistierenden Phasen ist ebenfalls eine Oberfl"ache.

Die Physik an der Oberfl"ache unterscheidet sich vielfach von der in
dem dazugeh"origen Volumen. Bei Festk"orpern kommt es vor, da"s das
Gitter an der Oberfl"ache einen andere Struktur hat. Im einfachsten
Fall hat die oberste Atomlage einen anderen Gitterabstand senkrecht
zur ihrer Ausbreitungsrichtung, als die Lagen tief im inneren des
Kristalls. Im Extremfall besitzt die Oberfl"ache eine ver"anderte
Gitterstruktur, indem sie eine "Uberstruktur ausbildet. Man spricht
von einer Rekonstruktion der Oberfl"ache, bei der die Gitterkonstante
parallel zur Oberfl"ache ein Vielfaches der Gitterkonstante im Volumen
ist. Zudem kann dieses in verschienen Richtungen ein unterschiedliches
Vielfaches sein.  Die Bandstruktur der Oberfl"ache unterscheidet sich
ebenfalls von der im Volumen. Somit "andern sich auch die
elektronischen Eigenschaften.

Die modernen experimentellen Methoden erm"oglichen es heute eine
atomare Oberfl"ache direkt zu beobachten. Speziell erw"ahnt seien hier
das Raster-Tunnelelektronen-Mikroskop und das
Raster-Kraft-Mikroskop. Ersteres erlaubt es die elektronische Struktur
der Oberfl"ache mithilfe eines Tunnelstroms, der durch eine winzige, im
idealen Fall einatomige Spitze flie"st, abzutasten. Das
Raster-Kraft-Mikroskop tastet die Oberfl"ache direkt mit einer Spitze
ab und gibt somit Aufschlu"s "uber den atomaren Aufbau.

In dieser Arbeit geht es darum einen der vielen Aspekte, welche eine
Oberfl"ache besitzt, etwas genauer zu beleuchten. Wir wollen deren
Rauhigkeit betrachten.  Es handelt sich dabei um eine
Gleichgewichtseigenschaft. Diese Eigenschaft hat auch
ihre Auswirkungen auf Nichtgleichgewichtsph"anomene. Als Beispiele
seien hier getriebene Grenzfl"achen, das Aufdampfen, auch bekannt
unter dem englischen Namen Molecular Beam Epitaxy (MBE), oder
allgemeiner das Kristallwachstum genannt.

Das hier betrachtete Modell ist nicht nur als Oberfl"achenmodell
interessant. Das reine Modell geh"ort zu einer ganzen Gruppe von
2-dimensionalen Modellen, die sich aufeinander abbilden lassen, wie das
$XY-Modell$ und das Coulombgas. Weiterhin gibt es mit diesem Modell
verwandte Probleme. Die Oberfl"ache beschreiben wir durch ein
skalares Feld, welchen durch einen in der Ebene liegenden Vektor
beschrieben wird. "Andern wir die Dimension dieses Parametervektors
z.B. auf den Wert eins und verwenden wir ein zweikomponentiges Felds,
so erhalten wir die Beschreibung eines gerichteten Polymeres. Mit einem
ganzen Ensemble solcher Linien kann man das Flu"sliniengitter eines Typ
II Supraleiters in der Abrikosov-Phase beschreiben. Als eine andere
Variation kann man ein Ensemble von Grenzfl"achen betrachten, womit
wir bei der Theorie der kommensurablen Systeme angekommen sind.

Die Arbeit gliedert sich folgenderma"sen: Zun"achst wird ein
Modell motiviert, mit dem man Oberfl"achen beschreiben kann. Es wird
auf einige bekannte Resultate von verwandten Modellen eingegangen. Im
darauffolgenden Kapitel wird das Modell mithilfe der 
Renormierungsgruppe untersucht. Im Rahmen einer
$\epsilon$-Entwichlungen werden Flu"sgleichungen gewonnen, aus denen
wir erste Resultate gewinnen. Danach folgt
eine Behandlung mittels eines Variationsverfahrens. Hierbei wird eine
Abwandlung der Renormierungsgruppe aus dem vorangehenden Kapitel
verwendet, die Drei-Parameter-Approximation.
Schlie"slich folgt ein Kapitel, in dem eine Methode vorgestellt wird,
mit der man auch in der urspr"unglichen Renormierungsgruppe zu 
einem Phasen"ubergang gelangen kann.

F"ur die Beschreibung einer Oberfl"ache hat man zwei
M"oglichkeiten. Die eine besteht darin, alles auf einer atomaren
L"angenskala zu betrachten. Hierbei wird die Position eines jeden
Atoms angegeben. Ein bekanntes Beispiel f"ur diese Art der
Beschreibung sind ist das Solid on Solid (SOS) Modell. Die andere
M"oglichkeit ist, "uber kleine Bereiche zu mitteln und somit zu einer
Kontinuumsbeschreibung "uberzugehen. Der zweite Zugang wird in dieser
Arbeit angewendet.

\chapter{Das Modell}
\section{Ein Oberfl"achenmodell am Beispiel des Ising-Modells}
Als Ausgangspunkt f"ur unsere Betrachtungen w"ahlen wir das
Ising-Modell auf einem Quadratgitter. Wir stellen uns vor, da"s auf
jedem Gitterpunkt ein Spin mit zwei Einstellm"oglichkeiten sitzt, der
Einfachheit halber $s= \pm 1$. Diese Spins k"onnen nun jeweils mit
ihren n"achsten Nachbarn wechselwirken. Die St"arke der Wechselwirkung
wird durch eine Kopplungskonstante $J$ beschrieben. Die
Hamiltonfunktion f"ur ein reines System, das hei"st ein System, wo die
Kopplungskonstanten zwischen den n"achsten Nachbarn alle gleich gro"s
sind, hat damit die folgenden Gestalt
\begin{equation*}
  \H_{\text{Ising}} = - \sum_{<i,j>} J s_is_j.
\end{equation*}
$<i,j>$ deutet an, da"s nur "uber n"achste Nachbarn summiert wird,
wobei jedes Paar nur einmal gez"ahlt wird. (Anderenfalls m"u"ste man
den Faktor $1/2$ erg"anzen.)

F"ur dieses System gibt es nun zwei F"alle mit je zwei entarteten
Grundzustandskonfigurationen, d.h. Konfigurationen mit minimaler
Energie $E$. Die $T=0$ Konfigurationen sind diese
Grundzustandskonfigurationen. Ist die Kopplungskonstante $J > 0$, so
richten sich die Spins parallel zueinander aus. Dieses ist der
ferromagnetische Fall. Der Grundzustand ist zweifach entartet, da das
System noch die M"oglichkeit hat, alle Spins nach oben oder nach unten
zeigen zu lassen. Bei $J < 0$, dem antiferromagnetischen Fall, sind
benachbarte Spins entgegengesetzt ausgerichtet. Das Gitter besteht aus
zwei Untergittern, in denen jeweils die Spins parallel ausgerichtet
sind. Auch hier gibt es zwei m"ogliche Konfigurationen, entweder
zeigen in dem einen Untergitter die Spins nach oben und in dem anderen
nach unten oder eben umgekehrt. Wir betrachten im folgenden den
Ferromagneten.

Vom Ising-Modell ist bekannt, da"s es in zwei und mehr Dimensionen
einen Phasen"ubergang besitzt \cite{LL}. Unterhalb der kritischen
Temperatur $T_c$ ist die Magnetisierung, das ist der thermisch
gemittelte Erwartungswert des Spins, ungleich null, oberhalb gleich
null. Der Phasen"ubergang ist von zweiter Ordnung, das hei"st die
Magnetisierung verschwindet kontinuierlich. Starten wir zum Beispiel
bei $T=0$ und erh"ohen die Temperatur, so findet ein in der
statistischen Physik immer wieder auftretender Proze"s statt. Das
Gleichgewicht ist durch das absolute Minimum der Freien Energie ${\cal
F} = E - TS$ bestimmt ($S$ ist die Entropie). Einige Spins drehen sich
um, wodurch sich die Energie des Systems erh"oht. Auf der anderen
Seite gewinnt das System aber an Entropie, da diese Spins an
beliebigen Pl"atzen im System liegen k"onnen. Mit wachsender
Temperatur bilden sich aus vereinzelten `verkehrt' stehenden Spins
kleine Bereiche, sogenannte Dom"anen. Der typische Durchmesser dieser
Dom"anen, welchen man als die {\em Korrelationsl"ange}
bezeichnet, w"achst, bis er bei $T_c$ die Systemgr"o"se
erreicht. Genauer mu"s man sagen, da es einen Phasen"ubergang nur im
thermodynamischen Limes gibt, da"s die Korrelationsl"ange bei $T_c$
divergiert. Unter dem thermodynamischen Limes versteht man, da"s man
die Systemgr"o"se gegen Unendlich gehen l"a"st, wobei man die Dichten
des Systems konstant h"alt.

Was uns nun interessiert sind die Begrenzungen dieser Dom"anen, die
sogenannten Do\-m"a\-nen\-w"an\-de. F"ur sie wollen wir eine geeignete
Hamiltonfunktion aufstellen und damit deren Eigenschaften
untersuchen. Zun"achst machen wir uns ein paar Gedanken "uber die
Energie einer solchen Wand. Die Wand befindet sich genau da, wo
die Spins entgegengesetzt stehen. Im Vergleich zu parallel
ausgerichteten Spins ist die Energie an solchen Stellen um $2 \cdot J$
gr"o"ser. Die Anzahl solcher {\em gebrochener}\/ Bindungen ist
proportional zur Fl"ache dieser Wand. Somit setzen wir die Energie der
Wand als eine Fl"achenenergiedichte multipliziert mit der Fl"ache der
Wand an.

\begin{figure}[htbp]
  \psfrag{x}{$\x$}
  \psfrag{z}{$z$}
  \psfrag{x1}{$x_1$}
  \psfrag{xD}{$x_D$}
  \psfrag{zx}{$z(\x)$}
  \centerline{\epsfig{figure=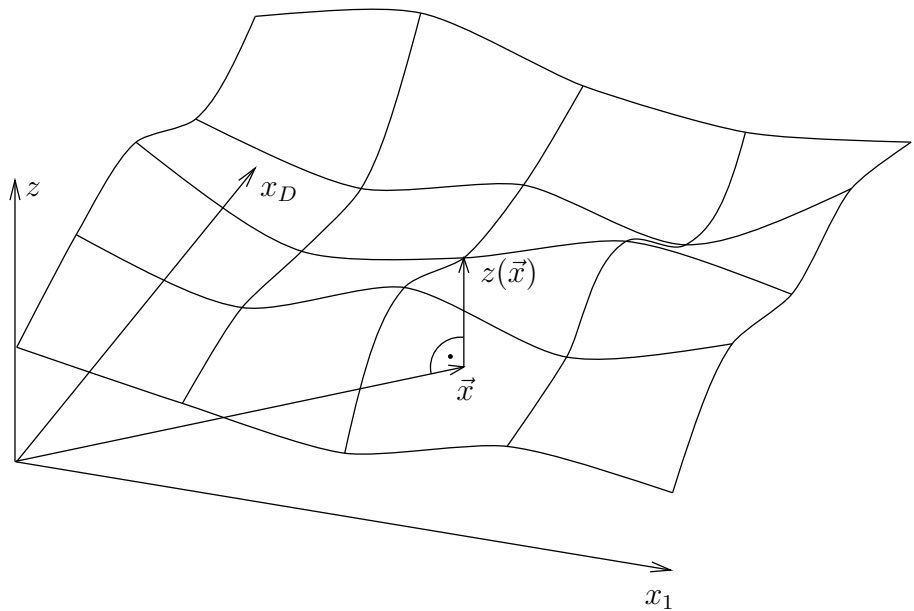,width=12cm}}
  \caption{\em Die Koordinaten einer gerichteten Fl"ache: der
  Parametervektor $\x$ spannt eine D-dimensionale Hyperebene auf; das
  Feld $z(\x)$ beschreibt das H"ohenprofil der Oberfl"ache "uber
  dieser Hyperebene}
  \label{fig:koordinaten}
\end{figure}

Was wir noch ben"otigen ist eine geeignete Parametrisierung der
Do\-m"a\-nen\-w"an\-de. In dem oben beschriebenen System ist dieses
mathematisch sehr kompliziert. Wir wollen uns daher auf eine einzige
Dom"anenwand beschr"anken. Eine solche Wand kann man im Ising-Modell
erzwingen, in dem man in einer Raumrichtung antiperiodische
Randbedingungen fordert. Diese ausgezeichnete Richtung wollen wir im
folgenden als die $z$-Richtung bezeichnen. Durch diese Randbedingung
mu"s das System im Verlauf der $z$-Richtung die Spinrichtung
wechseln. Im Grundzustand wird eine v"ollig ebene Wand
vorliegen. Eine solche Wand l"a"st sich nun einfach durch orthogonale
Koordinaten parametrisieren (Abbildung \ref{fig:koordinaten}). Die $D:=
d-1$ Koordinaten, entlang der Achsen senktrecht zur $z$-Richtung
bezeichnen wir mit dem Vektor $\x$, die entlang der $z$-Achse mit
$z$. Die zu untersuchende Fl"ache l"a"st sich nun als Feld $z(\x)$
beschreiben. Die Punkte der Fl"ache haben demnach die Koordinaten
$(\x, z(\x))$. Hierbei bezeichnen wir mit $d$ die Dimension des ganzen
Raumes und mit $D$ die Dimension der Fl"ache.  Eine Fl"ache, welche so
parametrisiert ist, bezeichnet man als eine gerichtete Fl"ache.

Den Preis, welchen man f"ur diese einfache Parametrisierung zu zahlen
hat, ist, da"s man mit ihr keine "Uberh"ange oder Einschl"usse
beschreiben kann. Die Frage ist nun, ob dieses unsere Betrachtungen
wesentlich beeinflu"st. Wir gehen zu einer vergr"oberten Beschreibung
(coarse graining) des Systems "uber, das hei"st, wir fassen einen
ausgedehnten Bereich der Oberfl"ache zusammen und summieren "uber die
Fluktuationen in diesem Bereich. Wenn wir diesen Bereich gro"s genug
w"ahlen, so werden die "Uberh"ange und Einschl"usse herausgemittelt.
Wir erreichen hierdurch  eine Kontinuumsbeschreibung der
Oberfl"ache. Der Einflu"s der Einschl"usse und "Uberh"ange schl"agt sich
nun in einer effektiven Energiedichte der Oberfl"ache nieder.
\begin{figure}[htbp]
  \centerline{\epsfig{figure=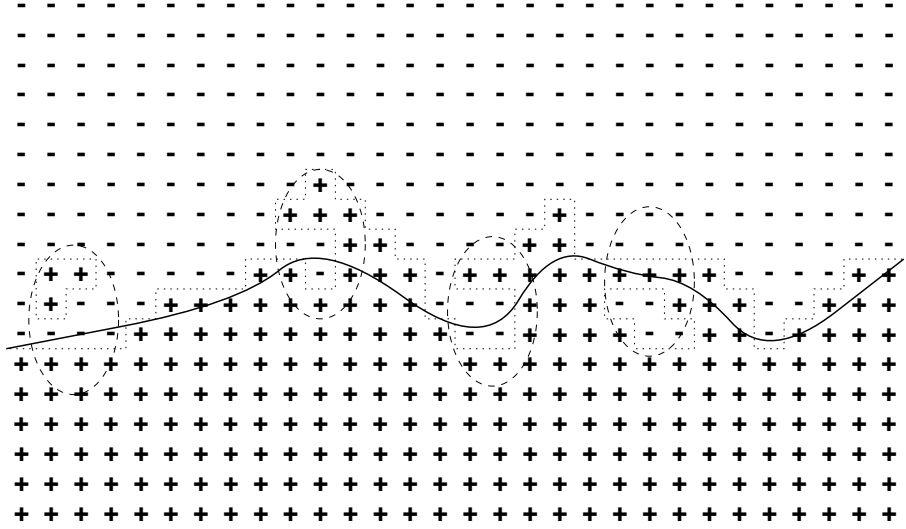,width=12cm}}
  \caption{\em Veranschaulichung des Vergr"oberungsprozesses: Auf
    jedem Gitterpunkt sitzt ein Spin, der entweder nach 
    oben $(+)$ oder nach unten $(-)$ zeigt. Die Grenze zwischen
    Bereichen $+$ und $-$ ist die zu untersuchende Wand. In dieser
    Konfiguration gibt es noch "Uberh"ange beziehungsweise
    Einschl"usse. Um diese zu beseitigen, mittelt man "uber kleinere
    aber hinreichend gro"se Bereiche (exemplarisch dargestellt durch
    die gestrichelten Ellipsen). Man endet bei einer vergr"oberten
    Beschreibung der Wand, die nun in etwa wie die durchgezogene
    wellige Linie aussieht.}
  \label{fig:coarsegraining}
\end{figure}

Wir k"onnen nun den ersten Beitrag zur Hamiltonfunktion aufschreiben.
\begin{displaymath}
  {\cal H}_{\text{el}} \propto \int d^Dx\ \Gamma\sqrt{1 + (\nabla z(\x))^2}
\end{displaymath}
$\Gamma$ ist die Oberfl"achensteifigkeit (oder
Oberfl"achenspannung). Das Integral "uber den Wurzelausdruck ist die
Fl"ache der Wand. F"ur kleine Gradienten kann man diesen Ausdruck nun
nach $(\nabla z(\x))^2$ entwickeln. In der niedrigsten Ordnung erh"alt man
\begin{equation}
  {\cal H}_{\text{el}} \propto \int d^Dx\ \frac \Gamma 2  (\nabla z(\x))^2.
\end{equation}
Hierbei haben wir den Konstanten Term weggelassen. Dieser beeinflu"st
nur die Grundzustandsenergie und hat somit keinen Einflu"s auf einen
Pha\-sen\-"uber\-gang. 

Bis hierher bezogen sich unsere Betrachtungen auf ein reines
System. Im folgenden wollen wir das System derart erweitern, da"s
wir Unordnung zulassen. Dabei wollen wir zwei verschiedene Arten von
Unordnung betrachten, Random Bond und Random Field Unordnung. 

Die erste Art der Unordnung kommt dadurch zustande, da"s im System an
einigen Stellen nichtmagnetische Atome sitzen. Im Ising-Modell
modelliert man eine solche Unordnung, indem man die Kopplungskonstante
$J$ durch eine Zufallsvariable ersetzt.
\begin{equation*}
  J = J_0 + \delta J
\end{equation*}
$J_0$ ist hier eine Konstante, etwa die Kopplungskonstante des reinen
Systems. $\delta J$ ist eine Zufallsvariable, welche von Ort zu Ort
einen anderen Wert besitzt. Ihr Mittelwert ist $\overline{\delta J} =
0$. "Ubersetzen wir dieses nun in die Formulierung unserer
Oberfl"achenhamiltonfunktion, so wird die Oberfl"achenspannung
$\Gamma$ eine Zufallsgr"o"se
\begin{displaymath}
  {\cal H} 
  \propto \int d^Dx\ (\Gamma + \delta\Gamma(\x,z(\x))) 
  \sqrt{1 + (\nabla z(\x))^2}.
\end{displaymath}
Wir entwickeln hier nun wieder bis zur ersten Ordnung in
$\delta\Gamma$ und $(\nabla z)^2$. 
\begin{displaymath}
  {\cal H} 
  \propto \int d^Dx\ \frac 12 \Gamma (\nabla z(\x))^2 +
  \delta\Gamma(\x,z(\x)). 
\end{displaymath}

Bei der zweiten Art der Unordnung, der Random Field Unordnung, befindet
sich das System in einem Zufallsfeld, einem Feld, welches an jedem Ort
im Raum einen zuf"alligen Wert annimmt. Dieses bedeutet, da"s jeder
Spin ein zuf"alliges `"au"seres' Feld sieht. Im Ising-Modell erh"alt
man hierdurch den zus"atzlichen Energiebeitrag $\sum_i H_is_i$, wobei
$H_i$ das Feld ist, welches der Spin auf dem i-ten Platz sieht. Die
Hamiltonfunktion der Wand erh"alt den Zusatz
\begin{equation*}
  \H_{\text{RF}} = \int d^Dx\ \int_0^{z(\x)} dz'\ H(\x,z') = \int
  d^Dx\ U(\x,z(\x)).
\end{equation*}

Bezeichnen wir nun im Random Bond Fall den Term $\delta\Gamma(\x,z)$
mit \-$U(\x,z)$, so haben wir in beiden F"allen dieselbe Gestalt der
Hamiltonfunktion. Zusammen mit dem elastischen Anteil lautet sie
\begin{equation}
  \H_{\text{ungeordnet}} = \int d^Dx\ \frac \Gamma 2 \left(\nabla
    z(x)\right)^2 + U(\x,z(\x)).
\end{equation}

Zwischen den beiden Arten der Unordnung unterscheidet man durch die
Art des Unordnungskorrelators. Die Random Bonds sind voneinander
unabh"angig, also unkorreliert. Ihr Korrelator ist somit lokal. Die
Random Fields sind ebenfalls voneinander unabh"angig verteilt, woraus
auch hier ein lokaler Korrelator f"ur das Feld $H(\x,z)$ folgt. Durch
die Integration in $z$-Richtung wird der Korrelator von $U(\x,z(\x))$
in dieser Richtung nichtlokal. Er verh"alt sich f"ur gro"se
Differenzen wie $-|z-z'|$.

Zusammengefa"st erh"alt man f"ur beide Unordnungen
\begin{equation*}
  \overline{U(\x,z)U(\x',z')} = \delta (\x -\x') R(z-z'),
\end{equation*}
wobei die Art der Unordnung durch die Funktion $R(z)$ unterschieden
wir. Sie hat in den beiden F"allen die Gestalt
\begin{equation*}
  \begin{aligned}[b]
    R_{\text{RB}}(z) &\propto \delta(z)\\
    R_{\text{RF}}(z) &\propto -|z|.
  \end{aligned}
\end{equation*}

Die Unordnung ver"andert das Verhalten des Systems dramatisch. Ist es
im Fall rein Thermischer Fluktuationen so, da"s die Energie der Wand
nur zu deren Fl"ache proportional ist, so kommt nun noch ein Beitrag
durch die Unordnung hinzu. Die Energie kann nun abgesenkt werden,
indem die Wand sich bevorzugt an solchen Stellen befindet, wo die
Kopplungskonstante zwischen den einzelnen Spins besonders klein
ist. Der Grundzustand ergibt sich nun aus der Konkurrenz der beiden
Energieterme, insbesondere mu"s er nicht mehr vollst"andig eben sein,
wie im reinen Fall.

Wir wollen nun in unsere Hamiltonfunktion noch einen dritten Term
ein\-f"u\-gen. Durch die Kontinuumsbeschreibung des Problems ist eine
Eigenschaft des Systems verlorengegangen, n"amlich die Tatsache, da"s
die Position der Wand immer zwischen den Spins liegt und somit
diskretisiert ist. Wir wollen dem nun Rechnung tragen, indem wir noch
ein zus"atzliches periodisches Potential in $z$-Richtung
einf"ugen. Die Wand m"ochte bevorzugt in einem Minimum dieses
Potentials sitzen. Somit f"uhren wir hiermit die Diskretisierung quasi
durch die `Hintert"ur' wieder ein. Wir haben also den additiven
Beitrag zur Energie
\begin{displaymath}
  {\cal H}_{\text{per}} = - \int d^Dx\ V_0 \cos \left( \frac {2\pi} a
    z \right) 
\end{displaymath}
$V_0$ ist die Amplitude des periodischen Potentials, $a$ der
`Gitterabstand' in $z$-Richtung. Hier ist es nun angebracht, eine
Bemerkung zu den Gitterkonstanten zu machen. Wir haben mit dem
Ising-Modell auf einem Quadratgitter begonnen. Die Gitterkonstante war
dort in allen Raumrichtungen gleichgro"s. Dieses mu"s nach dem
"Ubergang zum Kontinuumsmodell nicht mehr der Fall sein. Dabei wurden
die Fluktuationen der Wand "uber einen gewissen Bereich der Wand
herausgemittelt. Die Abmessungen dieses Bereiches in den
$\x$-Richtungen m"ussen nicht genau so gro"s sein wie die in die
$z$-Richtung. Wir wollen daher zwischen zwei Gitterkonstanten $\ap$
und $\as$ unterscheiden. Die erste ($\ap$) soll die Gitterkonstante
parallel zur Oberfl"ache, also die in den $\x$-Richtungen sein. Sie
definiert die kurzwellige Abschneidefrequenz $\Lambda = 1/\ap$ der
auftretenden Impulse (Ultraviolett-Cutoff). Die andere Gitterkonstante
$\as$ ist die Periodenweite des periodischen Potentials.

Die gesamte Hamiltonfunktion lautet somit
\begin{equation}
  \label{modell}
  {\cal H} = \int d^Dx\ \left\{ \frac \Gamma 2 \left( \nabla
      z(\x)\right)^2 -  V_0 \cos \left( \frac {2\pi} \as z \right) +
    U(\vec x,z(\x)) \right\}_.
\end{equation}
Dieses ist die Beschreibung der Oberfl"ache, welche den Betrachtungen
dieser Arbeit zugrunde liegt.

\section{Die Oberfl"achenrauhigkeit}
Der Hauptgesichtspunkt dieser Arbeit ist der Rauhigkeits"ubergang von
einer glatten zu einer rauhen Oberfl"ache. Um diesen untersuchen zu
k"onnen, ist es zun"achst einmal notwendig zu definieren, was man
unter der Rauhigkeit einer Fl"ache verstehen soll. Eine sinnvolle
Gr"o"se hierf"ur sind sicherlich die Fluktuationen der Grenzfl"ache um
ihren Mittelwert. Die Rauhigkeit einer Wand (beziehungsweise genauer
gesagt deren Quadrat) wird definiert als die mittleren quadratischen
Fluktuation der H"ohe $z(\vec x)$ (H"ohen-H"ohen-Korrelation).
\begin{equation}
  \label{rau}
  W(|\vec x - \vec x'|) = 
  \overline{\left<[z(\vec x) - z (\vec x')]^2\right>} ^{1/2}
\end{equation}
Wobei $\left<\dots\right>$ f"ur die thermische Mittelung und der
Querstrich $\overline{}$ f"ur die Unordnungsmittelung steht.

Konvergiert diese Gr"o"se f"ur gro"se Abst"ande $|\x-\x'|$ gegen einen
endlichen Wert, so handelt es sich um eine glatte
Oberfl"ache. Divergiert die Gr"o"se, so ist die Fl"ache rauh.  In der
rauhen Phase verh"alt sich die Rauhigkeit f"ur gro"se $L = |\x-\x'|$
wie
\begin{equation*}
  W(L) \propto L^\zeta.
\end{equation*}
Der kritische Exponent $\zeta$ hei"st {\em Rauhigkeitsexponent}.

\section{Die Behandlung von Systemen mit Unordnung} 

Ein zentraler Punkt bei der Behandlung von ungeordneten Systemen ist
die korrekte Handhabung der Unordnung. Insbesondere mu"s darauf
geachtet werden, da"s nur geeignete Gr"o"sen gemittelt werden. Diese
Gr"o"sen hei"sen selbstmittelnd. Die Begr"undung hierf"ur liegt darin,
da"s man zwischen dem wahrscheinlichsten Wert und dem Mittelwert einer
Gr"o"se unterscheiden mu"s. Der Mittelwert ist der Wert, den man bei
einer Mittelung der entsprechenden Gr"o"se erh"alt. F"uhrt man eine
Messung an einem realen System durch, so mi"st man jedoch den
wahrscheinlichsten Wert. Nun gibt es aber auch Zufallsgr"o"sen, bei
denen sich der wahrscheinlichste und der mittlere Wert wesentlich
unterscheiden. Dieses tritt bei stark asymmetrischen
Verteilungsfunktionen auf. Die Berechnung des Mittelwertes solcher
Gr"o"sen ist von sogenannten {\em seltenen Ereignissen} dominiert. Bei
einer Messung haben diese aber quasi keine Bedeutung, da sie eben
sehr selten sind.

Eine dieser Gr"o"sen ist die Freie Energie 
\begin{displaymath}
  F = - T \ln \text{Tr } e^{-\frac \H T} = -T \ln {\cal Z}.
\end{displaymath}
Hierbei ist die Boltzmannkonstante, wie auch im folgenden, gleich 1
gesetzt. 

Um diese Gr"o"se "uber die Unordnung zu mitteln, bedient man sich des
sogenannten {\it Replikatricks}. Zun"achst stellt man den Logarithmus
als einen Grenzwert dar. 
$$\ln x = \lim_{n \rightarrow 0} \frac 1 n (x^n - 1)$$

Die n-te Potenz der Zustandssumme ${\cal Z}^n$ wird nun als die Zustandssumme
eines n-fach vervielf"altigten Systems betrachtet. 
\begin{equation*}
  {\cal Z}^n = \left(\text{Tr}\ e^{-\frac \H T} \right)^n =
  \text{Tr}'\ e^{-\frac{\H_n} T}
\end{equation*}
Tr$'$ bedeutet hier, da"s die Spur "uber die Freiheitsgrade des
vervielf"altigten Systems gebildet wird. Dieses sind n mal so viele
Freiheitsgrade, wie im urspr"unglichen System.  $\H_n$ ist die
Hamiltonfunktion des gesammten oder {\em replizierten} Systems, also
aller Teilsysteme.  Die Teilsysteme, von denen jedes f"ur sich dem
urspr"unglichen System entspricht, sind zun"achst ungekoppelt. Die
Freiheitsgrade der Teilsysteme werden mit der Systemnummer des
Teilsystems indiziert $z_\alpha,\ \alpha=1..n$. Die Unordnung hingegen
ist unabh"angig von der Teilsystemnummer oder wie man auch sagt, vom
{\em Replikaindex}. Diese Zustandsumme des replizierten Systems wird
nun "uber die (gau"sverteilte) Unordnung gemittelt. Hierdurch entsteht
eine Kopplung zwischen den einzelnen Systemen.

Problematisch bei der Anwendung dieses Replikatricks ist, da"s hierbei
die Bildung des Grenzwertes $\lim_{n \rightarrow 0}$ und der
Unordnungsmittelung vertauscht werden. Es ist mathematisch nicht
sichergestellt, da"s man bei beiden M"oglichkeiten denselben Grenzwert
erh"alt. Dieses Problem k"onnen wir hier nicht weiter untersuchen und
erw"ahnen es hier nur der Vollst"andigkeit halber.

\section{Bekannte Resultate}

Das dieser Arbeit zugrundeliegende Problem wurde in dieser Form
bereits von Bouchaud und Georges mittels eines Variationsverfahrens
untersucht \cite{jpb+ag:prl92}. Sie approximieren dabei die ganze
replizierte Hamiltonfunktion
durch einen quadratischen Ausdruck und untersuchen das Problem mit
Hilfe der Replikasymmetriebrechung. F"ur Oberfl"achendimensionen $D$
gr"o"ser als zwei (Raumdimensionen $d$ gr"o"ser als drei) finden sie
drei verschiedene Phasen, eine glatte, eine glasartig glatte und eine
rauhe. In $D=2$ Dimensionen finden sie ebenfalls drei verschiedene
`Zust"ande' des Systems. Auf Grund von heuristischen Argumenten, die
einen Phasen"ubergang in 2 Oberfl"achendimensionen ausschlie"sen,
argumentieren sie, da"s die von ihnen gefundenen Grenzlinien
Crossoverlinien sein m"ussen. 

F"ur vereinfachte Modelle gibt es ebenfalls bekannte
Resultate. Solche Vereinfachungen bestehen darin, da"s man entweder das
periodische, das ungeordnete oder auch beide Potentiale wegl"a"st.

\subsection{Elastische Grenzfl"ache}
Das einfachste System besteht nur aus dem elastischen Anteil der
Hamiltonfunktion.
\begin{equation}
  \label{model:elastisch}
  \H_{\text{el}} = \int d^Dx\ \frac \Gamma 2 \left(\nabla z(\x)\right)^2
\end{equation}
Dieses Modell ist exakt l"osbar, da die Hamiltonfunktion quadratisch
im Feld $z$ ist. Die Rauhigkeit ergibt sich zu
\begin{equation*}
  W_{\text{el}}^2(x) = \frac T\Gamma \int \frac{d^Dq}{(2\pi)^D} \frac
  {1-e^{i\vec q \x}}{q^2}
\end{equation*}
Die Impulse $q$ im Integral sind nach unten begrenzt durch $1/L$ und
nach oben durch $1/\ap$. Es gibt hier drei m"ogliche F"alle zu
unterscheiden:

\fbox{$D > 2$} In diesem Fall nimmt das Integral f"ur $x,L \rightarrow
\infty$ einen endlichen Wert an. Die Grenzfl"ache ist also nach
unserer Definition glatt.

\fbox{$D = 2$} Dieses ist die sogenannte {\em marginale}
Dimension. Das bedeutet, die gr"o"ste Dimension bei der die
Oberfl"ache f"ur endliche Temperaturen noch rauh ist. Eine typische
Eigenschaft der marginalen Dimension ist, da"s die Rauhigkeit
logarithmisch mit dem Abstand $x$ w"achst und somit f"ur $x,L
\rightarrow \infty$ divergiert
\begin{equation*}
  W_{\text{el}}^2(x) \propto \ln \left(\frac x{\ap}\right)
\end{equation*}

\fbox{$D < 2$} Auch in diesem Fall ist die Oberfl"ache rauh. Im
Vergleich zum vorherigen Fall w"achst die Rauhigkeit allerdings nach
einem algebraischen Gesetz mit dem Abstand $x$. Dieses ist also
wesentlich schneller als im marginalen Fall.
\begin{equation*}
  W_{\text{el}}^2(x) \propto \frac T\Gamma L^{2-D}
\end{equation*}

Im zweiten und dritten Fall erh"alt man nur f"ur  $T = 0$ eine glatte
Oberfl"ache. F"ur endliche $T$ ist die Wand immer rauh. F"ur den
letzten Fall kann man den Rauhigkeitsexponenten $\zeta$ bestimmen. Er
betr"agt 
\begin{equation}
  \label{model:zeta_el}
  \zeta_{\text{el}} = \frac {2-D}2_.
\end{equation}

\subsection{Sinus-Gordon Modell}
Das n"achste Modell ist das Sinus-Gordon Problem. Hierbei setzt sich
die Hamiltonfunktion aus dem elastischen Term und dem periodischen
Potential zusammen. 
\begin{equation*}
  \H_{\text{el}} = \int d^Dx\ \left\{ \frac \Gamma 2 \left(\nabla
    z(\x)\right)^2 - V_0\cos \left(\frac {2\pi} \as z(\x)\right) \right\}
\end{equation*}

Dieses Modell ist nicht mehr exakt zu behandeln. Man mu"s geeignete
N"aherungen durchf"uhren, wie zum Beispiel eine Variationsrechnung
oder Renormierungsgruppenrechnung
\cite{chuiweeks:76,chuiweeks:78,yi:zpb78,sak_et_al:82}. Auch bei
diesem Modell ist die marginale Dimension $D=2$. Im Fall $D>2$ ist die
Wand somit glatt. F"ur $D=2$ "andert sich das Verhalten derart, da"s
es nun einen durch die Temperatur getriebenen Phasen"ubergang
gibt. Unterhalb einer kritischen Temperatur $T_c$ ist die Wand glatt,
w"ahrend sie sich oberhalb in einer rauhen Phase befindet. Die Art des
Phasen"ubergangs f"allt in die Universalit"atsklasse des Kosterlitz
Thoules "Ubergang.

\subsection{Elastische Oberfl"ache in Ungeordneten Systemen}
Ein weiteres bekanntes Modell ist die elastische Grenzfl"ache in einem
ungeordneten Medium \cite{dsf:86}, beschrieben durch die
Hamiltonfunktion
\begin{equation*}
  \H = \int d^Dx\ \left\{ \frac \Gamma 2 \left(\nabla
      z(\x)\right)^2 + U(\x,z(\x)) \right\}_.
\end{equation*}

F"ur Dimensionen $D<4$ ist dieses System durch einen $T=0$ Fixpunkt
bestimmt. Das System ist durch die Unordnung bedingt selbst bei
$T=0$ rauh. Der kritische Exponent $\zeta$ l"a"st sich mit Hilfe einer
funktionalen Renormierungsgruppenrechnung bestimmen. Man erh"alt in
der 1-loop Approximation:
\begin{equation}
  \begin{aligned}[b]
    \zeta_{\text{RF}} &= \frac {4-D}3\\
    \zeta_{\text{RB}} &\approx (4-D)\cdot 0.2083
  \end{aligned}
\end{equation}
Der Random Bond Exponent ist ein numerisches Ergebnis.

In einer Dimension $(D=1)$ l"a"st sich das Problem f"ur Random Bond
Unordnung mit Hilfe von Transfermatrizen exakt l"osen. Man erh"alt 
\begin{equation}
  \zeta = \frac 23,
\end{equation}
was nicht ganz mit dem numerischen Resultat von Fisher
"ubereinstimmt. Hierbei ist zu beachten, da"s die Ergebnisse von
Fisher im Rahmen der f"uhrenden Ordnung einer $\epsilon$ Entwicklung
gewonnen wurden $(\epsilon = 4-D)$.

\section{Skalenargumente}
Einen guten "Uberblick erh"alt man bereits, wenn man das
Skalenverhalten der auftretenden Gr"o"sen betrachtet. Dabei reskaliert
man die auftretenden L"angen und Impulse derart, da"s man zu
gr"o"seren L"angenskalen "ubergeht (wir interessieren uns
haupts"achlich f"ur das Verhalten auf gro"sen L"angen). Durch die
Forderung der Invarianz der Hamiltonfunktion oder die Equillibierung
von Energiebeitr"agen erh"alt man die kritischen Exponenten.

Als einf"uhrendes Beispiel wollen wir das Modell der elastischen
Grenzfl"ache (\ref{model:elastisch}) betrachten. Wir betrachten, wie
sich die Energie einer Ausbeulung der Oberfl"ache mit deren Gr"o"se
"andert.  Die L"angen parallel zur Oberfl"ache skalieren wie
$L$. Somit liefert das Integral $\int d^Dx$ einen Beitrag $L^D$. $z$
ist auf dieser L"angenskala die Rauhigkeit $W(L)$. Der Gradient
$\nabla$ liefert eine inverse L"angenpotenz $L^{-1}$. Somit skaliert
die gesamte Hamiltonfunktion wie $\Gamma L^{D-2}W^2$. Diese Energie
mu"s von der thermischen Energie aufgebracht werden, wir setzen den
Ausdruck also gleich $T$ und l"osen nach $W$ auf.
\begin{equation*}
  W \approx \left(\frac T \Gamma \right)^{1/2}L^{(2-D)/2}
\end{equation*}
Wir erhalten denselben Rauhigkeitsexponenten wie in
(\ref{model:zeta_el}) $\zeta = (2-D)/2$. Auch der Vorfaktor
$(T/\Gamma)^{1/2}$ ist derselbe.

\subsection{Das Imry Ma Argument}
Wir wollen nun eine elastische Grenzfl"ache in einem ungeordneten
Medium betrachten und die Energiebeitr"age des elastischen Term und der
Unordnung equillibrieren. Dieses f"uhrt zum {\em Imry Ma
Argument}.

Bei diesem Argument betrachtet man die Beitr"age zur Energie bei
$T=0$, die man gewinnt und verliert, wenn man einen Bereich mit dem
Durchmesser $L$ einer glatten Oberfl"ache um eine Distanz $W$ in
$z$-Richtung verformt. Es entsteht dabei ein Energieverlust durch den
elastischen Term. Dieser ist genau so gro"s wie im Fall ohne Unordnung,
skaliert also wie $\Gamma L^{D-2}W^2$.

Die Energie, die notwendig ist, um diese Verformung zu
realisieren, kann nur durch die Unordnung aufgebracht werden.
Das ungeordnete Potential verschwindet im Mittel. Wir m"ussen also die
Fluktuationen derselben betrachten. Weiterhin mu"s zwischen Random
Bond und Random Field Unordnung unterschieden werden. Im ersten Fall
gewinnt die Oberfl"ache an Energie, indem sie m"oglichst viele Minima
des ungeordneten Potentials erreicht. Die Fluktuationen sind also
proportional zur Fl"ache der Ausbeulung und skalieren somit wie
$L^D$. 

Der Mechanismus, durch den die Oberfl"ache im Random Field Fall
Energie gewinnen kann, funktioniert anders. Gebrochene Bindungen haben
nur dann eine Auswirkung auf die Energie, wenn sie an der Oberfl"ache
liegen. Im Volumen spielt es keine Rolle, ob die an dieser Bindung
liegenden Spins beide nach oben zeigen oder durch die verschobene
Oberfl"ache beide nach unten. Die Random Field Unordnung wirkt auf die
Einstellung eines jeden Spins und nicht auf die relative Einstellung
zweier benachbarter Spins. Wird die Oberfl"ache verr"uckt, so kommt
die Energie"anderung dadurch zustande, da"s die Spins in dem von der
Oberfl"ache durchwanderten Volumen ihre Einstellung umgedreht
haben. Die Energiefluktuationen sind also proportional zu diesem
Volumen und skalieren somit wie $L^{D}W$.

Wir equillibrieren nun die beiden Energiebeitr"age. Dabei ist zu
beachten, da"s der Energiebeitrag der Unordnung die Wurzel aus den
Fluktuationen ist. Nachdem wir nach der Rauhigkeit $W$ aufgel"ost
haben, k"onnen wir den Rauhigkeitsexponenten f"ur beide Arten der
Unordnung ablesen
\begin{equation}
  \label{model:imry_ma}
  \begin{aligned}[b]
    \zeta_{\text{RF}} &= \frac {4-D}3\\
    \zeta_{\text{RB}} &= \frac {4-D}4
  \end{aligned}
\end{equation}

F"ur den Random Bond Fall ist die Rauhigkeit geringer als f"ur den
Random Field Fall. Der Grund hierf"ur ist, da"s die Random Bonds nur an
der Oberfl"ache angreifen, die Random Fields jedoch am Volumen. 

\subsection{Das Flory Argument}
Eine weitere Methode, den Rauhigkeitsexponenten zu bestimmen, ist das
Flory Argument. Hierbei skaliert man die L"angen und fordert, da"s die
reduzierte Hamiltonfunktion $\H/T$ invariant bleibt, sowie da"s
der elastische und der ungeordnete Anteil gleich skalieren. Durch
diese Forderungen wird der kritische Exponent $\zeta$ festgelegt.
Die L"angen $x$ und $z$ skalieren wie $\tilde x = x/b$ und $\tilde z =
z/b^\zeta$. Mit der Tilde $\tilde {}$ werden reskalierte Gr"o"sen
bezeichnet. Die Unordnung skaliert wie die Wurzel des Korrelator. 
\begin{equation*}
  \begin{aligned}[b]
    \tilde U_{\text{RF}}(\tilde x,\tilde z) &=
    b^{(D-\zeta)/2}U_{\text{RF}}(\x,z) \\
    \tilde U_{\text{RB}}(\tilde x,\tilde z) &=
    b^{(D+\zeta)/2}U_{\text{RB}}(\x,z)
  \end{aligned}
\end{equation*}
Somit ergibt sich
\begin{equation}
  \begin{aligned}[b]
    \zeta_{\text{RF}} &= \frac {4-D}3\\
    \zeta_{\text{RB}} &= \frac {4-D}5.\\
  \end{aligned}
\end{equation}

Bei Random Field Unordnung stimmen die Exponenten, die man aus Imry Ma
und Flory Argument gewinnt, "uberein. Anders ist dieses bei Random
Bond Unordnung. In einer Oberfl"achendimension haben wir die
M"oglichkeit, die beiden Rauhigkeitsexponenten
$\zeta_{\text{Flory}}=3/5$ und $\zeta_{\text{Imry Ma}} = 3/4$ mit dem
bekannten exakten Resultat $\zeta_{\text{exakt}}=2/3$ zu
vergleichen. Es stellt sich heraus, da"s
\begin{equation*}
  \zeta_{\text{Flory}} < \zeta_{\text{exakt}} < \zeta_{\text{Imry Ma}}.
\end{equation*}

Das Imry Ma Argument "ubersch"atzt den Exponenten, w"ahrend die
Reskalierung der Hamiltonfunktion ihn untersch"atzt. Eine Begr"undung
hierf"ur zeigen exakte Grundzustandsuntersuchungen f"ur $D=1$ dimensionale
Systeme \cite{mk:jap87}. In Abbildung (\ref{mod:rbrf}) sind die
optimalen Pfade (dieses sind $D=1$ dimensionale Oberfl"achen) zwischen
einem gemeinsamen Anfangspunkt und einem variierenden Endpunkt f"ur
Random Bond und Random Field Unordnung dargestellt. 
Die Grafiken zeigen einen markanten Unterschied zwischen Random Field
und Random Bond Unordnung. Im Random Field Fall ``teilen'' sich
benachbarte Pfade wesentlich fr"uher als im Random Bond
Fall. Benachbarte Pfade enthalten bei Random Bond Unordnung viele
gemeinsame Segmente. Daher mu"s man die Annahme "uber den
Energiegewinn durch die Unordnung, welche dem Imry Ma Argument
zugrunde liegt, als eine obere Grenze betrachten. Der
Rauhigkeitsexponent $\zeta_{\text{Imry Ma}}$ ist damit ebenfalls eine
obere Schranke.
\newpage
$$\quad$$
\vfill
\centerline{(a)}
\vfill
\begin{figure}[h]
  \centerline{(b)}
  \caption{\em Die optimalen Grundzust"ande von Benachbarten
    Grenzfl"achen bei Random Field (a) und Random Bond (b) Unordnung
    in einer Oberfl"achendimension: Es werden Fl"achen betrachtet, die
    einen gemeinsamen Anfangspunkt besitzen (Spitze des Dreiecks) und
    deren Endpunkte auf der Basis des Dreiecks liegen. Deutlich
    zusehen ist, da"s im Random Field Fall benachbarte Fl"achen eine
    wesentlich geringere Anzahl gemeinsamer Segmente haben als im
    Random Bond Fall. (Abbildung aus \cite{mk:jap87})}
  \label{mod:rbrf}
\end{figure}
\label{frg}

In diesem Kapitel wird die Oberfl"ache mit Hilfe der funktionalen
Renormierung untersucht. Sie unterscheidet sich von der ``normalen''
beziehungsweise, genauer gesagt, von der Parameter Renormierung
dadurch, da"s nicht nur die Kopplungskonstanten in der
Hamiltonfunktion renormiert werden, sondern da"s sich die ganze
funktionale Form der Beitr"age zur Hamiltonfunktion "andert.
Auf der formalen Seite haben wir bei der Funktionalen Renormierung
anstatt von gew"ohnlichen Differentialgleichungen f"ur die
Kopplungskonstanten, wie bei der Parameter Renormierung, nun partielle
Differentialgleichungen f"ur die funktionalen Anteile der
Hamiltonfunktion, wie beispielsweise des periodischen Potentials.

Die funktionale Renormierungsgruppentransformation besteht aus drei
Teilschritten.  Zuerst wird das Feld $z(\x)$ in einen langsam und
einen schnellfluktuierenden Anteil aufgespaltet. Dieses entspricht der
Zerlegung des Impulsraumes in eine Kugelschale mit den Impulsen
$\Lambda / b < |\vec q\,| < \Lambda$ und in eine Kugel mit $|\vec
q\,| < \Lambda/b$ zerlegt. $\Lambda$ ist hierbei der obere Cutoff f"ur
die Impulse, der durch das zugrundeliegende Gitter zustande
kommt: $\Lambda \propto 1/{\ap}$. $b > 1$ ist ein dimensionsloser
Parameter. "Uber die Impulse in der Kugelschale wird integriert. Die
dabei entstehenden Terme mu"s man zu geeigneten Gr"o"sen in der
verbleibenden Hamiltonfunktion, die nur noch von den `kleinen'
Impulsen abh"angt, addieren. Dadurch entstehen effektive Gr"o"sen. Man
erh"alt so aus der urspr"unglichen Hamiltonfunktion mit einem Cutoff
von $\Lambda$ eine, bei der sich der Cutoff auf $\Lambda/b$ reduziert
hat. Als dritten Schritt skaliert man alle L"angen und Impulse so,
da"s der urspr"ungliche Cutoff wiederhergestellt wird.

Wir betrachten eine D-dimensionale Grenzfl"ache in einem
D+1-dimensiona\-len Raum. Die Fl"ache wird durch einen D-dimensionalen
Vektor $\vec x$ parametrisiert. Die Auslenkung der Fl"ache sei durch
das einkomponentige Feld $z(\vec x)$ gegeben. 
Wir beschreiben die Fl"ache mit der  folgenden reduzierten
Hamiltonfunktion:
\begin{equation}
  {\cal H} = \int d^Dx \frac 1T \left\{\frac 1{2}\Gamma (\nabla
      z(\vec x))^2 - V_0\cos\left(\frac {2\pi}{\as} z(\vec x)\right) +
      U(z(\vec x),\vec x)\right\}_.
\end{equation}
Reduzierte Hamiltonfunktion bedeutet, da"s diese durch die Temperatur
$T$ dividiert wurde. Im weiteren Verlauf dieses Kapitels soll der
Begriff Hamiltonfunktion immer die reduzierte Hamiltonfunktion meinen.

Der erste Term beschreibt die elastische Energie der Oberfl"ache,
der zweite ist ein periodisches Potential, welches die diskrete
Struktur des Gitters simuliert. Der letzte Term ist das
Unordnungspotential, welches gau"sverteilt ist und durch die beiden
Momente
\begin{eqnarray*}
  \overline{V(z,\vec x)} &=& 0 \\ 
  \overline{V(z,\vec x)V(z',\vec x')} &=& \delta(\vec x - \vec x')
  R(z-z')
\end{eqnarray*}
charakterisiert wird. Alle h"oheren Momente lassen sich durch das Wick
Theorem auf diese beiden zur"uckf"uhren.  Der Querstrich "uber den
Gr"o"sen symbolisiert hierbei die Unordnungsmittelung. Mit der
Funktion $R(z)$ wird die Art der Unordnung modelliert. F"ur die beiden
in dieser Arbeit betrachteten Unordnungformen hat sie die Gestalt
\begin{equation*}
  R(z) \propto \left\{
      \begin{array}{cl}
        -|z| & \text{ random field}\\
        \delta(z) & \text{ random bond}
      \end{array}
    \right._.
\end{equation*}

Die Unordnungsmittelung behandeln wir mit Hilfe des Replikatricks. Die
replizierte Hamiltonfunktion hat folgende Form
\begin{multline}
  \label{repH}
  \quad{\cal H}^n = \int d^Dx \left\{  \sum_\alpha \frac 1T \left[ \frac
      \Gamma {2} \left(\nabla z^\alpha(\vec x)\right)^2 -
      V_0\cos\left(\frac {2\pi}{\as} z^\alpha(\vec x) \right) \right]
  \right. \\ \left. - \frac 1 {2T^2}  
    \sum_{\alpha,\beta}R\left(z^\alpha(\vec x) -z^\beta(\vec
      x)\right)\right\} 
\end{multline}
Die griechischen Buchstaben $\alpha$ und $\beta$ sind die
Replikaindizes, sie laufen von 1 bis $n$. Im Weiteren wird die
Abk"urzung $V(z) := V_0 \cos\left(2 \pi z/{\as} \right)$
verwendet.

\section{Skalierung}
\label{skal}
Zun"achst wollen wir uns einen "Uberblick verschaffen und das
Verhalten der einzelnen Anteile in der Hamiltonfunktion bei Reskalierung der
auftretenden L"angen betrachten. Wir ersetzen die Gr"o"sen $\x$ und
$z$ durch $\tx$ und $\tilde z$ mit 
\begin{displaymath}
  \x = e^l \tx, \qquad z = e^{\zeta l} \tilde z.
\end{displaymath}
Die Auslenkung $z$ der Fl"ache mu"s mit dem Rauhigkeitsexponenten $\zeta$
skaliert werden, da sie wie die Rauhigkeit skaliert.  Hier und
im folgenden deutet die Tilde $\tilde{\ }$ an, da"s es sich um
reskalierte Gr"o"sen handelt. Diese setzen wir in die replizierte
Hamiltonfunktion ein. Wir fordern nun, da"s die Kopplungskonstante
$\Gamma$ des elastischen Terms unver"andert bleibt und erhalten somit
das Skalenverhalten der Temperatur.
\begin{equation*}
    \int d^D\tilde xe^{Dl} \frac \Gamma {2T} \left(\frac
      {\text{d}}{\text{d}\tx}\tilde ze^{(\zeta -1)l}\right)^2 = 
    \int d^D\tilde x \frac \Gamma {2\tilde T}
    \left(\frac{\text{d}}{\text{d}\tx}\tilde z\right)^2
\end{equation*}
\begin{equation*}
    \Rightarrow \quad
    \tilde T = e^{(2-D-2\zeta)l}T
\end{equation*}
Mit diesem $\tilde T$ bestimmen wir nun ganz analog $\tilde R(\tilde
z)$ und $\tilde V(\tilde z)$
\begin{eqnarray*}
  \tilde R(\tilde z) &=& e^{(4-D-4\zeta)l}R(e^{\zeta l}\tilde z)\\
  \tilde V(\tilde z) &=& e^{(2-2\zeta)l} V(e^{\zeta l}\tilde z)
\end{eqnarray*}

Die hieraus folgenden Flu"sgleichungen lauten
\begin{align}
  \frac {\partial\tilde T}{\partial l}=& (2-D-2\zeta)\tilde T
  \tag{\theequation .a}\\ 
  \frac {\partial\tilde R}{\partial l} =& (4-D-4\zeta)\tilde R + \zeta
  \tilde z \tilde R'\tag{\theequation .b}\label{ren:scalR}\\ 
  \frac {\partial\tilde V}{\partial l} =& (2-2\zeta)\tilde V+ \zeta
  \tilde z \tilde V'\tag{\theequation .c} 
\end{align}
\addtocounter{equation}{1}%
An diesen Flu"sgleichungen kann man bereits zwei wesentliche Dinge
ablesen. Die Temperatur $T$ flie"st f"ur $D > 2-2\zeta$ gegen null. In
dem f"ur uns interessanten Bereich $D \geq 2$ haben wir es also mit
einem $T=0$ Fixpunkt zu tun. Wir m"ussen also bei der folgenden
Funktionalen Renormierung nur solche Terme mitnehmen, die auch im
Grenzwert $T \rightarrow 0$ einen Beitrag liefern, die anderen fallen
weg. An der Flu"sgleichung (\ref{ren:scalR}) f"ur $\tilde R$ sieht
man, da"s die obere kritische Dimension 4 ist. D.h. oberhalb von 4
Dimensionen ist die Unordnung irrelevant. Die Grenzfl"ache bleibt also
wie f"ur $D>2$ Dimensionen im reinen Fall glatt.

\section{Funktionale Renormierungsgruppe}
Wir beginnen die Funktionale Renormierung in der "ublichen Art und
Weise mit der Zerlegung des Feldes $z(\vec x)$ in den lang- und
kurzwelligen Anteil.
\begin{eqnarray*}
  z(\vec x) &=& z^<(\vec x) + z^>(\vec x)\\
  z^>(\x) &=& \frac 1 {\sqrt{Vol}} \left.\sum_q\right.^> z_{\vec q}\
  e^{i\vec q \vec x}\\
  z^<(\vec x) &=& \frac 1 {\sqrt{Vol}} \left.\sum_q\right.^< z_{\vec q}\
  e^{i\vec q \vec x}
\end{eqnarray*}

$\sum_q^>$ ist dabei eine Abk"urzung f"ur die Summe "uber Impulse mit
$\Lambda/b < |\vec q\,| <\Lambda$, $\sum_q^<$ f"ur die "uber Impulse mit
$|\vec q\,| < \Lambda/b $.

Der elastische Term ist diagonal in diesen beiden Teilfeldern. Das
periodische Potential sowie der Unordungsanteil m"ussen noch nach
diesen Feldern zerlegt werden. Beide Terme werden nach den
kurzwelligen Feldern bis zur zweiten Ordnung entwickelt.
\begin{equation*}\begin{split}
  V(z) = V\left(z^< + z^>\right) = V\left(z^<\right) +
  V'\left(z^<\right)z^> + \frac 1 2 V''\left(z^<\right)z^{>2} 
  + O(z^{>3})
\end{split}
\end{equation*}
Der Strich $'$ symbolisiert die Ableitungen nach $z$. Bei der
Funktion $R$ ist noch darauf zu achten, da"s sie nicht diagonal in den
Replikaindizes ist. Zur Vereinfachung f"uhren wir die Abk"urzung
$z^{\alpha \beta} := z^\alpha - z^\beta$ ein.
\begin{multline*}
  R\left(\zab\right) = R\left(\zab^< + \zab^>\right) \nonumber \\
  = R\left(\zab^<\right) + R'\left(\zab^<\right)\zab^> + \frac 1 2
  R''\left(\zab^<\right)\zab^{>2} 
  + O\left(\zab^{>3}\right)
\end{multline*}
Bei der Summation "uber die Replikaindizes kann man die Symmetrie der
Funktion $R$
\begin{eqnarray*}
  R(z) = R(-z) \quad\Rightarrow\quad R'(z)=-R'(-z)
  \quad\Rightarrow\quad R''(z) = R''(-z)
\end{eqnarray*}
sowie $\zab = -z^{\beta\alpha}$ ausnutzen. Man erh"alt somit
\begin{multline*}
  \begin{split}
  &\quad\sum_{\alpha, \beta}
  R'\left(\zab^<\right)\left(z^{\alpha>}-z^{\beta>}\right) =
  \sum_{\alpha, \beta} 2 \cdot R'\left(\zab^<\right)z^{\alpha>} \\
  &\quad\sum_{\alpha, \beta}
  R''\left(\zab^<\right)\left(z^{\alpha>}-z^{\beta>}\right)^2 
  \end{split}\\
   = -2
  \sum_{\alpha, \beta} \underbrace{\left( R''\left(\zab^<\right) -
  \sum_\gamma
  R''\left(z^{\alpha\gamma<}\right)\delta^{\alpha\beta}\right)}_{=: 
  Q^{\alpha\beta} = Q^{\beta\alpha}}z^{\alpha>} z^{\beta>}\ .\nonumber\quad
\end{multline*}

Wir  f"uhren folgend Fouriertransformationen ein:
\begin{eqnarray*}
  V^{'\alpha}(\vec q) &:=& \int d^Dx\ \frac 1 {\sqrt{Vol}}\
  e^{i\vec q \vec x}\ V'\left(z^{\alpha<}(\vec x)\right)\\
  V^{''\alpha}(\vec q) &:=& \int d^Dx\ \frac 1 {Vol}\ e^{i\vec q \vec
  x}\ V''\left(z^{\alpha<}(\vec x)\right) \\
  R^{'\alpha\beta} (\vec q) &:=& \int d^Dx\ \frac 1
  {\sqrt{Vol}}\ e^{i\vec q \vec x}\
  R'\left(z^{\alpha\beta<}(x)\right)\\ 
  Q^{\alpha\beta} (\vec q) &:=& \int d^Dx\ \frac 1 {Vol}\ e^{i\vec q
  \vec x}\ Q^{\alpha\beta} 
\end{eqnarray*}

und erhalten die Zerlegung der Hamiltonfunktion
\begin{equation*}
  \begin{split}
    \H =& \H\left(z^{\alpha\beta <}\right) + \H_0\left(z^{\alpha>}\right) +
    \H_1\left(\zab^<,\zab^>\right) + O(z^{>3})\\ 
    \intertext{mit}
    \H_0\left(z^{\alpha >}\right) =& 
    \sum_\alpha \Sum{\vec q}\frac
    \Gamma {2T} q^2 z_{\vec q}^\alpha  z_{-\vec q}^\alpha\\
    \H_1\left(\zab^<,\zab^>\right) =& \\
    - \frac 1 T &\sum_\alpha \left(\Sum{\vec q} V^{'\alpha} (\vec
      q)z_{\vec q}^\alpha    \Sum{\vec q,\vec q\,'}   
      + \frac 1 2  V^{''\alpha} (\vec q+\vec q\,')z_{\vec q}^\alpha
    z_{\vec q\,'}^\alpha \right) \\
     - \frac 1 {2 T^2} &\sum_{\alpha,\beta} \left(
      \Sum{\vec q}  R^{'\alpha\beta}(\vec q) z_{\vec q}^\alpha 
      - \Sum{\vec q,\vec q\,'} Q^{\alpha\beta}(\vec q + \vec q\,')z_{\vec 
        q}^\alpha z_{\vec q\,'}^\beta  \right)_.
  \end{split}
\end{equation*}
$\H(\zab^<)$ hat dieselbe funktionale Gestalt wie die urspr"ungliche
Hamiltonfunktion. Der einzige Unterschied zwischen beiden ist, da"s
diese nur noch von den kleinen Impulsen abh"angt. $H_0(z^{\alpha >})$
ist eine quadratische Hamiltonfunktion, die nur gro"se Impulse
enth"alt und zudem diagonal in den Replikaindizes
ist. $\H_1(\zab^<,\zab^>)$ h"angt von allen erlaubten Impulsen
ab. Somit werden hierdurch die Anteile mit kleinen und gro"sen
Impulsen gekoppelt. Durch die Renormierungsgruppentransformation
entstehen aus diesem Teil der Hamiltonfunktion die Korrekturen zu
$\H(\zab^<)$.

Hiermit k"onnen wir nun die kurzwelligen Freiheitsgrade
ausintegrieren, das hei"st wir bilden die Zustandssumme, wobei wir nur
"uber die Feldanteile mit den gro"sen Impulsen integrieren. Das
Resultat interpretieren wir als $\exp(-\H_{\text{renormiert}}(\zab^<))$.

\begin{eqnarray*}
  \left.\prod_{\vec q}\right.^> \int_{-\infty}^\infty dz_{\vec q}\
  e^{-{\cal H}^n} &=& e^{-{\cal H}(z^<)}
  \left<e^{-{\cal H}_1(z^<,z^>)}\right>_{{\cal H}_0(z^>)}Z_0^>
  \nonumber \\ 
  &=& e^{-{\cal H}(z^<)} Z_0^> \exp\left<e^{-{\cal H}_1(z^<,z^>)} -
  1 \right>_{{\cal H}_0(z^>),C} \nonumber \\
  &=& e^{-{\cal H}(z^<)} Z_0^> e^{-\delta {\cal H}}
\end{eqnarray*}

Um die Kumulanten zu berechnen, ben"otigt man die Erwartungswerte der
verschiedenen Potenzen von $z_{\vec q}$. Da die Hamiltonfunktion
${\cal H}_0$ quadratisch und diagonal in den Impulsen $\vec q$ ist,
kann man das {\em Wick Theorem} anwenden.  Die Erwartungswerte mit
einer ungeraden Anzahl von $z$ verschwinden. Die mit einer geraden
werden auf den quadratischen Erwartungwert zur"uckgef"uhrt, wobei
\begin{equation*}
  \left<z_{\vec q}^\alpha z_{\vec q\,'}^\beta\right>_{\H_0(z^>)} =
  \delta^{\alpha,\beta}\delta_{\vec q,- \vec q\,'} \frac T {\Gamma q^2}
\end{equation*}
ist.

In erster Ordnung erh"alt man:
\begin{eqnarray*}
  \delta{\cal H}^{(1)} &=& -\frac 1 {2T} \Sum{\vec q,\vec q\,'}
    \sum_{\alpha} V^{''\alpha} (\vec q+\vec q\,') \delta_{\vec q,-\vec
    q\,'} \frac T {\Gamma q^2} \nonumber \\ 
  && + \frac 1 {2T^2} \Sum{\vec q,\vec q\,'} \sum_{\alpha,\beta}
    Q^{\alpha\beta}(\vec q + \vec q\,') 
    \delta^{\alpha,\beta}\delta_
    {\vec q,- \vec q\,'} \frac T {\Gamma
      q^2} \nonumber \\
  &=& \Sum{\vec q}\sum_\alpha \left( \frac 1 2 V^{''\alpha}(0) + \frac
    1 {2 T} Q^{\alpha\alpha}(0)\right)
  \frac 1 {\Gamma q^2} 
\end{eqnarray*}

Die hier auftretende Summe "uber die Impulse $q$ n"ahern wir durch ein
Integral:
\begin{eqnarray*}
  \Sum{\vec q} \frac 1 {q^2} \approx Vol \int^> \frac {d^Dq}{(2\pi)^D}
  \frac 1 {q^2} &=& Vol\ K_D \int^\Lambda_{\Lambda/b} dq\ q^{D-3}
  \nonumber \\ &=& Vol\ K_D \frac 1 {D-2} \Lambda^{D-2}(1- b^{2-D})
\end{eqnarray*}

$K_D$ ist dabei die Oberfl"ache der D-dimensionalen Einheitskugel
dividiert durch $(2\pi)^D$. Dieser Term zeigt f"ur $D > 2$ und $b
\rightarrow \infty$ keine Divergenz. Somit k"onnen die Beitr"age der
ersten Ordnung vernachl"assigt werden.

Als n"achstes wird die zweite Ordnung betrachtet:
\begin{multline*}
  \delta {\cal H}^{(2)} = \frac 12 \left< \left[ - \frac 1 T \sum_\alpha
      \left(\Sum{\vec q} V^{'\alpha} (\vec q\,)z_{\vec q}^\alpha
        \Sum{\vec q,\vec q\,'} + \frac 1 2 V^{''\alpha} (\vec q+\vec
        q\,')z_{\vec q}^\alpha z_{\vec q\,'}^\alpha \right) \right. \right.
  \nonumber \\
  \left.\left. - \frac
 1 {2 T^2} \sum_{\alpha,\beta} \left( \Sum{\vec q} \tilde
 R^{'\alpha\beta}(\vec q\,) z_{\vec q}^\alpha - \Sum{\vec
 q,\vec q\,'} Q^{\alpha\beta}(\vec q + \vec q\,')z_{\vec q}^\alpha
 z_{\vec q\,'}^\beta \right) \right]^2 \right>_{{\cal H}_0(z^>),C} \nonumber
\end{multline*}

Durch Ausmultiplizieren des Quadrats $[\dots]^2$ entstehen 10
verschiedene  Summanden, die als Kandidaten f"ur Korrekturen in Frage
kommen. Vier sind von der Ordnung $z^3$ und fallen somit bei der
Mittelung weg. Drei sind quadratisch in $z$:
\begin{multline*}\qquad
  \frac 1 {2T^2} \sum_{\alpha\alpha'}\Sum{\vec q \vec q\,'}
  V^{'\alpha}(\vec q) V^{'\alpha'}(\vec
  q\,')\delta^{\alpha,\alpha'}\delta_{\vec q,- \vec q\,'} \frac T {\Gamma
    q^2} \\=
  \frac 1 {2T} \sum_\alpha\Sum{\vec q}V^{'\alpha}(\vec
  q) V^{'\alpha}(-\vec q) \frac 1 {\Gamma q^2} \qquad\nonumber
\end{multline*}
\begin{multline*}\qquad
  \frac 1 {8T^4} \sum_{\alpha,\beta \atop \alpha'\beta'}\Sum{\vec q,
    \vec q\,' } R^{'\alpha\beta}(\q\,) \tilde
  R^{'\alpha'\beta'}(\q\,') \delta^{\alpha,\alpha'} \delta_{\q,\q\,'}\frac
  T {\Gamma q^2} \\=
  \frac 1 {8T^3} \sum_{\alpha,\beta \atop \beta'}\Sum{\vec q} \tilde
  R^{'\alpha\beta}(\q\,) 
  R^{'\alpha\beta'}(-\q\,)\delta_{\q,-\q\,'}\frac 1 {\Gamma q^2}
  \qquad\nonumber 
\end{multline*}
\begin{multline*}\qquad
  \frac 1 {2T^3}\sum_{\alpha,\beta \atop \alpha'} \Sum{\q,\q\,'}
  R^{'\alpha\beta}(\q\,)V^{'\alpha'}(\q\,')\delta^{\alpha,\alpha'}
  \delta_{\q,-\q\,'}\frac T{\Gamma q^2} \\=
  \frac 1 {2T^2}\sum_{\alpha,\beta}\Sum{\q}
  R^{'\alpha\beta}(\q\,)V^{'\alpha}(-\q\,)\frac T{\Gamma q^2}
  \qquad\nonumber 
\end{multline*}
Bei diesen drei  Beitr"agen  taucht wieder $\sum_{\vec q}^>1/q^2$
auf, so da"s sie keine Divergenz f"ur $b \rightarrow \infty$
zeigen. Sie werden also vernachl"assigt.

Der Term, der quadratisch in $V''$ ist, ist von der Ordnung $T^0$. Wie
wir bereits weiter oben gesehen haben, handelt es sich bei diesem
Problem um einen $T=0$ Fixpunkt. Die Terme, die einen Beitrag liefern,
m"ussen also von derselben Ordnung in $T$ sein, wie die Gr"o"sen in
der urspr"unglichen replizierten Hamiltonfunktion. Der Term des
periodischen Potentials ist von der Ordnung $T^{-1}$. Somit wird der
in $V''$ quadratische Term vernachl"assigt.

Der quadratische Term in $Q$ sieht folgenderma"sen aus:
\begin{multline}
  \label{frg:int2}
  \frac 1 {8 T^4} \sum_{\alpha,\beta \atop \alpha'\beta'}\Sum{\vec q,
    \vec q\,' \atop \vec k,\vec k'} Q^{\alpha\beta}(\vec q + \vec
  q\,') Q^{\alpha', \beta'}(\vec k + \vec k') \left<\left(z_{\vec
        q}^\alpha z_{\vec q\,'}^{\beta}\right) \cdot \left(z_{\vec
        k}^\alpha z_{\vec k\,'}^{\beta'}\right)\right>_{\H_0(z^>),C} \\
  = \frac 1 {4 \Gamma^2 T^2}\sum_{\alpha,\beta}\Sum{\vec q, \vec q\,'}
  Q^{\alpha\beta}(\vec q + \vec q\,')Q^{\alpha\beta}(-\vec
  q - \vec q\,') \frac 1 {q^2} \frac 1 {q^{'2}} 
\end{multline}

Als erstes ersetzen wir die Summe "uber $\vec q$ und $\vec q\,'$ durch
entsprechende Integrale und f"uhren $\vec k = \vec q + \vec q\,'$ ein:
\begin{multline*}
  \Sum{\vec q, \vec q\,'} Q^{\alpha\beta}(\vec q
  + \vec q\,')Q^{\alpha\beta}(-\vec q - \vec q\,') \frac 1 {q^2} \frac
  1 {q^{'2}}  \\ 
  = \text{Vol}^2 \int^{o} \frac {d^Dk}{(2\pi)^D}
  \int^> \frac {d^Dq}{(2\pi)^D} Q^{\alpha\beta}(\vec k) \tilde
  Q^{\alpha\beta}(-\vec k) \frac 1 {q^2}\frac 1 {(\vec k + \vec
    q\,)^2}
\end{multline*}

Der Kringel $^o$ am Integral zeigt an, da"s $k$ so gew"ahlt sein soll,
da"s $\vec q$ und $\vec q\,'$ in der Kugelschale mit $\Lambda/b < |\vec
q\,| < \Lambda$ liegen. Wir nehmen an, da"s die $Q$'s nur einen Beitrag
f"ur kleine $k$ liefern. Dann n"ahern wir $ 1 /(\vec k + \vec
  q)^2 \approx 1 / {q^2}$. Somit vereinfacht sich das
Integral zu
\begin{equation*}
  \begin{aligned}[b]
  \text{Vol}^2\int^{o} & \frac {d^Dk}{(2\pi)^D}  \int^>   \frac 
  {d^Dq}{(2\pi)^D} \tilde 
  Q^{\alpha\beta}(\vec k)  Q^{\alpha\beta}(-\vec k) \frac 1
  {q^4} \\
    &= \text{Vol}^2
    \int^{o} \frac {d^Dk}{(2\pi)^D} Q^{\alpha\beta}(\vec
    k)  Q^{\alpha\beta}(-\vec k) 
    \frac {K_D}{\epsilon} \Lambda^{-\epsilon} (b^\epsilon -1)\\
    &= \int d^Dx Q^{\alpha\beta}Q^{\alpha\beta} \frac
    {K_D}{\epsilon} \Lambda^{-\epsilon} (b^\epsilon -1) .
    \end{aligned}
\end{equation*}
Dieser Term zeigt eine Divergenz f"ur $D < 4$ und $b \rightarrow
\infty$. Wir setzen nun die $Q^{\alpha\beta}$ ein:
\begin{multline*}
  \frac 1 { 4 \Gamma^2 T^2} \int d^Dx
    \sum_{\alpha, \beta}  \Bigl\{
    R^{''2}\left(\zab^<\right) - 2 R^{''}(0)R^{''}\left(\zab^<\right)
    \\
    + R^{''}\left(\zab^<\right)\sum_{\gamma}R^{''}
    \left(\left.z^{\alpha\gamma}\right.^<\right) \frac
    {K_D}{\epsilon} \Lambda^{-\epsilon} (b^\epsilon -1) 
  \Bigr\} \nonumber
\end{multline*}
Hier haben wir drei verschiedene Beitr"age. Alle drei sind von der
Ordnung $T^{-2}$. Die ersten beiden enthalten zwei Replikaindizes,
somit liefern sie einen Beitrag zur Renormierung von $R$. Der letzte
Term h"angt von drei Replikaindizes ab. Er w"urde also eine
`Drei-Replika-Kopplung', die in der urspr"unglichen Hamiltonfunktion
nicht vorhanden ist, renormieren. Es bleibt die Frage, ob dieser Term
eine drei Replika Kopplung generieren kann. Wir betrachten dazu, wie
eine Kopplung zwischen drei Replikas in der replizierten
Hamiltonfunktion zustande kommen k"onnte. Die Mischterme entstehen
durch die Unordnungsmittelung. Um einen Term mit drei Replikaindizes
zu generieren, m"u"ste die Unordnungsverteilung noch nichtgau"ssche
Anteile enthalten. Au"ser dem quadratischen m"u"ste es noch andere
Momente der Unordnungsverteilung geben, die sich nicht durch das
quadratisch Moment ausdr"ucken lassen.  Eine hieraus entstehende
Kopplung von drei Replikas w"are aber von der Ordnung
$T^{-3}$\cite{dsf:85}. Somit ist dieser Term f"ur einen $T=0$ Fixpunkt
irrelevant, da er von der Ordnung $T^{-2}$ ist. Folglich
vernachl"assigen wir ihn.

Es bleibt noch ein Mischterm zwischen dem ungeordneten und
periodischen Potential zu betrachten. 
\begin{equation*}
  \begin{aligned}[b]
    -\frac 1 {2T^3}&\sum_{\alpha,\beta \atop \alpha'} \Sum{\q, \q\,'
      \atop \vec k, \vec k'} Q^{\alpha\beta}(\q + \q\,') 
    V^{''\alpha'}(\vec k+\vec k') \delta^{\alpha,\alpha'}
    \delta^{\beta,\alpha'} \delta_{\q,-\vec k} \delta_{\q\,',-\vec
      k'}\frac T {\Gamma q^2} \frac T {\Gamma q^{'2}}\\
    &= -\frac 1 {2\Gamma^2T}\sum_\alpha \Sum{\q,\q\,'}
      Q^{\alpha\alpha}(\q + \q\,')V^{''\alpha}(-\q-\q\,')\frac 1
      {q^2}{q^{'2}}\\ 
    &= -\frac {\text{Vol}^2}{2\Gamma^2T} \sum_{\alpha}\int^o \frac
      {d^Dk}{(2\pi)^D}  
    Q^{\alpha\alpha}(\vec k)V^{''\alpha}(-\vec k)
    \frac {K_D}\epsilon \Lambda^{-\epsilon}(b^\epsilon -1)
  \end{aligned}
\end{equation*}
Hierbei wurde wieder $\vec k = \q + \q\,'$ eingef"uhrt und die Annahme
gemacht, da"s das Integral nur Beitr"age f"ur kleine $\vec k$
liefert. In der Ortsdarstellung haben wir nun
\begin{eqnarray*}
  -\frac 1 {2\Gamma^2T} \sum_\alpha \int d^Dx\ \left[ R''(0)
    -\sum_\beta R''\left(\zab^<\right)\right] V''\left(z^{\alpha
      <}\right) \frac {K_D}\epsilon \Lambda^{-\epsilon}(b^\epsilon -1)
    \nonumber. \\ \quad
  \end{eqnarray*}
Der erste Summand h"angt von einem Replikaindex ab. Er ist von der
Ordnung $T^{-1}$. Somit handelt es sich um eine Korrektur zu $V$. Der
zweite enth"alt eine Summe "uber zwei Replikaindizes, er k"onnte also
einen Beitrag zu $R$ liefern. Dazu hat er allerdings die falsche
Ordnung in $T$. Er f"allt also weg.

Damit haben wir alle Terme in dieser Ordnung betrachtet. Wir erhalten
insgesamt die effektiven Gr"o"sen
\begin{eqnarray}
  R_{eff} &=& R(y) + \frac 1 2 \frac {K_D}{\Gamma^2}\frac 1 \epsilon 
  \Lambda^{-\epsilon}(b^\epsilon -1
  )\left[R^{''2}(z)-2R''(z)R''(0)\right]\\ 
  V_{eff} &=& V(z) - \frac 1 2 \frac {K_D}{\Gamma^2}\frac 1 \epsilon 
  \Lambda^{-\epsilon}(b^\epsilon -1 ) V''(z) R''(0).
\end{eqnarray}

F"ur die Temperatur $T$ ergeben sich in dieser Ordnung der Entwicklung
keine Korrekturen. Die replizierte Hamiltonfunktion hat nun wieder die
Gestalt von (\ref{repH}), wobei das periodische Potential und der
Unordnungskorrelator durch die effektiven Gr"o"sen ersetzt worden
sind. Der obere Cutoff hat sich von $\Lambda$ auf $\Lambda/b$
reduziert.

Wir reskalieren die auftretenden L"angen wie in Abschnitt \ref{skal}
und stellen somit den urspr"unglichen Cutoff wieder her. Mit einem
infinitesimal gew"ahlten $l= \ln (b)$ erhalten wir folgendes
Flu"sgleichungssystem:

\begin{align}
  \label{frg:fluss}
    \frac {\partial\tilde T}{\partial l} &= (2 - D  - 2\zeta) \tilde T
    \tag{\theequation .a} \\ 
    \frac {\partial\tilde R}{\partial l} &= (\epsilon - 4\zeta)\tilde
    R + \zeta \tilde z\tilde R ' + \frac{K_D\Lambda^{-\epsilon}}{2\Gamma^2}
    \left[\tilde R^{''2}(\tilde z)-2\tilde R''(\tilde z) \tilde
    R''(0)\right] \tag{\theequation .b}\\   
    \frac {\partial\tilde V}{\partial l} &= (2 - 2\zeta) \tilde V +
    \zeta \tilde z \tilde V ' - \frac 1 2 \frac{K_D}{\Gamma^2}
    \Lambda^{-\epsilon}\tilde V''(\tilde z)\tilde R''(0)\tag{\theequation .c}
\end{align}
\addtocounter{equation}{1}%

Die ersten beiden Gleichungen entsprechen genau denen von D.S. Fisher
\cite{dsf:86}. Neu hinzugekommen ist die dritte Gleichung f"ur das
periodische Potential. Eine erweiterte Form dieser Gleichung taucht
bei J. Kierfeld beim Problem gekoppelter Schichtsupraleiter auf
\cite{jkf:93}.  Wir definieren nun noch
\begin{equation*}
  \tilde D(\tilde z) := \frac {K_D\Lambda^{-\epsilon}}{\Gamma^2} \tilde
  R(\tilde z),
\end{equation*}
um uns von den physikalisch unbedeutenden Konstanten in den
Flu"sgleichungen zu befreien und erhalten
\begin{align}
  \label{frg:fluss2}
  \frac {\partial\tilde T}{\partial l} &= (2 - D  - 2\zeta) \tilde T
  \tag{\theequation .a}\\  
  \frac {\partial\tilde D}{\partial l} &= (\epsilon - 4\zeta)\tilde
  D + \zeta \tilde z\tilde D ' + \frac 1 2 \left[\tilde
    D^{''2}(\tilde z)-2\tilde D''(\tilde z) \tilde D''(0)\right]
  \tag{\theequation .b}\\   
  \frac {\partial\tilde V}{\partial l} &= (2 - 2\zeta) \tilde V +
  \zeta \tilde z \tilde V ' - \frac 1 2 \tilde V''(\tilde z)\tilde
  D''(0) . \tag{\theequation .c}
\end{align}
\addtocounter{equation}{1}%

\section{Fixpunkte der Flu"sgleichungen}
\label{frg:fp}
Um die verschiedenen Phasen zu bestimmen, ist es n"otig, die Fixpunkte
der Flu"sgleichungen zu bestimmen. Besonders einfach ist dieses f"ur
die Temperatur. $T=0$ ist hier ein Fixpunkt. Da die zugeh"orige
Flu"sgleichung eine lineare, gew"ohnliche Differentialgleichung
erster Ordnung ist, kann man sie auch vollst"andig l"osen
\begin{equation*}
  \tilde T(l) = T(l=0) e^{(2-D-2\zeta)\cdot l}.
\end{equation*}
Somit wird die Temperatur f"ur gro"se $l$ und Dimensionen 
$D > 2-2\zeta$, was gro"sen L"angenskalen entspricht, gegen null
getrieben. $T=0$ entspricht einem stabilen Fixpunkt. Er
beschreibt somit eine Phase und keinen Phasen"ubergang. Dieses
Resultat haben wir schon bei der Herleitung der Flu"sgleichung f"ur
das periodische Potential und den Unordnungskorrelator verwendet.

Schwieriger ist es bei den anderen beiden Flu"sgleichungen, da es sich
hier um partielle Differentialgleichungen handelt. Weiterhin m"ussen
wir noch anstelle eines Anfangswertes, wie bei der Temperatur,
Randbedingungen, die durch die urspr"unglichen, unrenormierten
Gr"o"sen gegeben sind, ber"ucksichtigen.  F"ur das periodische
Potential $V$ ist die Randbedingung, da"s die Funktion periodisch
ist. Die Randbedingung des Unordnungskorrelator ist bei
den beiden Unordnungsarten unterschiedlich. Im Random Field Fall ist die
Randbedingung die Asymptotik des Korrelators f"ur gro"se $z$, 
$R(z) \propto -|z|$. Im Random Bond Fall hingegen die Tatsache, da"s die
Korrelationen kurzreichweitig sind.

Wir betrachten zun"achst die Flu"sgleichung f"ur den
Unordnungskorrelator. Dieses ist genau dieselbe Flu"sgleichung, die
D.S. Fisher in seiner Arbeit zu einem "ahnlichen Problem, n"amlich
dem ohne Periodischen Potential, gefunden hat \cite{dsf:86}. Wir
erhalten also f"ur die beiden Unordnungsarten dieselben Fixpunkte wie
er.

Zun"achst betrachten wir den Random Field Fall. Setzen wir hier das
asymptotische Verhalten des Unordnungskorrelators ein, so ist damit
der Rauhigkeitsexponent $\zeta$ bereits festgelegt.
\begin{equation}
  \zeta = \frac \epsilon 3
\end{equation}
Mit diesem $\zeta$ haben wir nun die Gleichung 
\begin{equation*}
  -\frac \epsilon 3 \tilde D(\tilde z) + \frac \epsilon 3 \tilde z \tilde
  D'(\tilde z) + \frac 12 \left[\tilde D^{''2}(\tilde z)-2\tilde
  D''(\tilde z) \tilde D''(0)\right] = 0  
\end{equation*}
zu l"osen. Leiten wir diese Gleichung nun einmal nach $\tilde z$ ab
und dividieren sie danach durch $\tilde D''(\tilde z)$, so erhalten wir
eine elementar integrable Gleichung. Diese ausgef"uhrt ergibt nun eine
implizite Gleichung f"ur die zweite Ableitung der Fixpunktfunktion
$\tilde D^{*{''}}(\tilde z)$
\begin{equation}
    \tilde\Delta^*\ln\left( -
      \frac {\tilde D^{*{''}}(\tilde z)}{\tilde\Delta^*}\right) +
    \tilde\Delta^* +  \tilde D^{*{''}}(\tilde z) 
    = - \frac
    \epsilon 6 \tilde z^2, 
\end{equation}
wobei $\tilde\Delta^*:=- \tilde D^{*{''}}(0)$ gesetzt wurde.
$\tilde\Delta^*$ ist durch die obige Gleichung noch nicht bestimmt und
parametrisiert somit eine ganze Familie von Fixpunktfunktionen, die
sich durch einen unterschiedlichen L"angenma"sstab in $z$-Richtung
unterscheiden. Eine nat"urliche Wahl f"ur $\tilde\Delta^*$ ist
\begin{equation}
  \tilde\Delta^* = \epsilon.
\end{equation}

Im Falle von Random Bond Unordnung kann man die Fixpunktfunktion nur
numerisch bestimmen. F"ur gro"se $\tilde z$ verh"alt sich hier der Korrelator
am Fixpunkt wie
\begin{equation}
  \tilde D^*(\tilde z) \propto \epsilon \tilde z^{-5+ \frac \epsilon \zeta}e^{- \frac
    {\tilde z^2}{2\epsilon}}.
\end{equation}
Die numerische Berechnung liefert den Rauhigkeitsexponenten
\begin{equation}
  \zeta \approx 0.2083\epsilon
\end{equation}

Als letztes m"ussen wir nun die Fixpunktfunktion f"ur das periodische
Potential finden. Die Unordnung geht hier durch den Faktor $\tilde
D^{''}(0)$ ein. Setzen wir hier $\tilde V \equiv 0$, so stellen wir fest,
da"s dieses ein Fixpunkt der Flu"sgleichung ist. Wir m"ussen jetzt
noch die Stabilit"atseigenschaften dieses Fixpunktes untersuchen.  Die
Reskalierung in $z$-Richtung bringt den schwierigen Term $\zeta z
\tilde V'(z)$ in die Flu"sgleichung hinein. Um diesen zu vermeiden,
betrachten wir das zwar renormierte aber {\em nicht} reskalierte
effektive Potential.
\begin{displaymath}
  V_{eff}(z) = V(z) - \frac 1 2 \frac 1 \epsilon V''(z)D''(0)(b^\epsilon-1)
\end{displaymath}
Mit $q_0 := \Lambda/b$ erhalten wir somit die Flu"sgleichung 
\begin{displaymath}
  \frac {\partial V}{\partial q_0} = \frac 1 2 
  \frac 1 \epsilon V''(z)D''(0) q_0^{-\epsilon -1}
\end{displaymath}
Da es sich hierbei um Gr"o"sen handelt, die nicht reskaliert wurden,
der gefundene Fixpunkt $\tilde D$ aber eine reskalierte Gr"o"se ist,
mu"s die Skalierung noch r"uckg"angig gemacht werden.
\begin{eqnarray*}
  \tilde D(\tilde z) &=& b^{\epsilon - 4\zeta}D(b^\zeta \tilde z)\\
  \Rightarrow \tilde D''(\tilde z) &=& b^{\epsilon - 2\zeta}D''(b^\zeta
  \tilde z)\\
  \Rightarrow D''(0)&=& b^{-\epsilon+2\zeta}\tilde D''(0) =
  -b^{-\epsilon+2\zeta}\epsilon
\end{eqnarray*}
Weiterhin sehen wir, da"s sich der Cosinus des periodischen Potentials
durch die zweite Ableitung von $V$ nach $z$ reproduziert. Wir
nehmen nun an, da"s durch die Renormierung nur der Vorfaktor $V_0$ des
periodischen Potentials beeinflu"st wird. Somit erhalten wir hier eine
Flu"sgleichung f"ur eben diesen Vorfaktor $V_0$:
\begin{displaymath}
  \frac {\text d V_0}{\text d q_0} = \frac 1 2 \Lambda^{2\zeta}
  \left(\frac{2\pi}{\as}\right)^2\epsilon 
  q_0^{-2\zeta-1} V_0
\end{displaymath}
Wir erhalten also eine lineare Differentialgleichung der Ordnung
1. Diese k"onnen wir nun durch Trennung der Variablen l"osen. Wie
integrieren die resultierende Gleichung $q_0=1/L$ bis $q_0=\Lambda$
und erhalten
\begin{equation}
  V_0(L)=V_0(\ap) \cdot \exp \left({-
  \left(\frac{2\pi}{\as}\right)^2 \frac {\epsilon}{4\zeta}
  \left((\Lambda L)^{2\zeta}-1\right)}\right).
\end{equation}
Wir sehen, da"s das periodische Potential auf gro"sen L"angenskalen
immer irrelevant wird. Das bedeutet, da"s unsere Grenzfl"ache immer
rauh ist. Die hier vorgestellte Rechnung liefert also keinen
Phasen"ubergang zwischen einer glatten und einer rauhen Phase. Die Art
der vorhandenen Unordnung geht hier nur "uber den
Rauhigkeitsexponenten $\zeta$ ein. Somit verschwindet das periodische
Potential f"ur den Random Field Fall schneller als f"ur den Random
Bond Fall. Dieses vertr"agt sich mit der Vorstellung, da"s die Random
Field Unordnung eine st"arkere Unordnung als die Random Bond Unordnung
ist. Bei der Random Bond Unordnung spielen nur die Defekte in einer
relativ kleinen Umgebung der Grenzfl"ache eine Rolle. Die
zus"atzlichen Energiebeitr"age der Zufallsfelder erh"alt man, indem
man die Fl"ache von $z = \pm \infty$ in das System hineinschiebt. Es
handelt sich hierbei also um eine Art Volumeneffekt. Dies spiegelt
sich auch in der Korrelationsfunktion wieder, f"ur Random Fields ist
sie langreichweitig, f"ur Random Bonds hingegen lokal.

\chapter{Variationsrechnung}

In diesem Kapitel wollen wir den Einflu"s des periodischen Potentials
mit Hilfe der Variationsrechnung untersuchen. 
Bei diesem Verfahren wird die urspr"ungliche Hamiltonfunktion durch
eine einfachere ersetzt. Diese vereinfachte Hamiltonfunktion enth"alt
noch Parameter, die zu bestimmen sind. Dieses geschieht durch die
Minimierung eines geeigneten {\em Variations-Funktionals} (Anhang
\ref{a:funktional}) 
\begin{equation*} 
  {\cal F}^{\text{var}} := F_0 + \left< \H -  \H_0 \right>_{\H_0}.
\end{equation*}
Hierbei ist $\H$ die urspr"ungliche und $\H_0$ die
Variations-Hamiltonfunktion. $F_0$ ist die Freie Energie der
Variations-Hamiltonfunktion:
$$F_0 = - T \ln \text{Tr}\ e^{-\frac {\H_0} T}$$ 

Auch hier rechnen wir wieder in $\epsilon = 4 -D$ Dimensionen. 
Wir ersetzen das periodische Potential durch ein quadratisches, im
folgenden Masseterm genannt. Der zu variierende Parameter ist die
Kopplungskonstante dieses Terms, die Masse m. Wir haben also
\begin{eqnarray}
  {\cal H}  &=& \int d^Dx\ \left\{ \frac
    \Gamma 2 \left(\nabla z(\vec x)\right)^2 -
    V_0\cos\left(\frac {2\pi}{\as} z(\vec x) \right)
    + U(z(\x),\x) \right\} \\
  {\cal H}_0  &=& \int d^Dx\ \left\{ \frac
    \Gamma {2} \left(\nabla z(\vec x)\right)^2 +
    \frac 1 2 m^2 z^2(\x)  + U(z(\x),\x) \right\}
\end{eqnarray}
Hierbei handelt es sich wieder um  replizierte
Hamiltonfunktionen. Den Index $n$ haben wir hier und im folgenden
weggelassen.

Wir wollen nun das Variations-Funktional minimieren. Dazu berechnen
wir
\begin{equation*}
  \frac {\partial \F^{\text{var}}}{\partial m} 
  = \frac{\partial}{\partial m} \left\{-T \ln \Z_0 + \frac 1 {\Z_0}
    \text {Tr} \left[(\H-\H_0) e^{-\frac {\H_0}T} \right]\right\},
\end{equation*}
wobei $\Z_0 = \text{Tr} \exp(-\H_0/T)$ die Zustandssumme bez"uglich
der Hamiltonfunktion $\H_0$ ist. Diesen Ausdruck gleich null. Wir
benutzen die Tatsache, da"s nur $\H_0$ und $\Z_0$ von $m$ abh"angen:
\begin{equation*}
  \begin{aligned}
    0 = &\left\{-T \frac 1 {\Z_0} - \frac 1 {\Z_0^2}\text {Tr} \left[(\H-\H_0)
        e^{-\frac {\H_0}T} \right] \right\} \frac {\partial \Z_0}{\partial
      m} \\
    &+ \frac 1{\Z_0} \text{Tr} \left[ - \frac {\partial \H_0}{\partial
        m}\left(1+ \frac 1 T (\H-\H_0)\right) e^{-\frac {\H_0}T}\right]
  \end{aligned}
\end{equation*}

Wir berechnen weiter
\begin{equation*}
  \frac {\partial \Z_0}{\partial m} = \frac \partial{\partial m}
  \text{Tr}\ e^{-\frac {\H_0}T} = \text{Tr}\left( -\frac 1 T \frac{\partial
      \H_0}{\partial m} e^{-\frac {\H_0}T} \right)
  = -\frac {\Z_0} T \left<\frac{\partial \H_0}{\partial m}\right>_{\H_0}
\end{equation*}
und erhalten somit
\begin{equation*}
  \begin{aligned}
    0 =& \left<\frac{\partial \H_0}{\partial m}\right>_{\H_0} +
    \frac 1 T \left<\H - \H_0\right>_{\H_0} \left<\frac{\partial
        \H_0}{\partial m}\right>_{\H_0} \\ 
    &- \left<\frac{\partial \H_0}{\partial m}\right>_{\H_0} - \frac 1
    T \left<(\H - \H_0)\frac{\partial \H_0}{\partial m}\right>_{\H_0}\\
    =& \frac 1 T \left<\H -\H_0\right>_{\H_0}\left<\frac{\partial
        \H_0}{\partial m}\right>_{\H_0}  - \frac 1 T \left<(\H -
      \H_0)\frac{\partial \H_0}{\partial m}\right>_{\H_0}.
  \end{aligned}
\end{equation*}

Diese k"onnen wir noch als Kumulanten-Erwartungswert
zusammenfassen und erhalten somit nach der Multiplikation mit $T$ die
Selbstkonsistenzgleichung
\begin{equation}
0 = \left<\left(\H -\H_0 \right)\cdot \frac {\partial \H_0}{\partial
    m}\right>_{\H_0,C}.
\end{equation}

Wir m"ussen nun diese Kumulante berechnen. Wir nehmen dazu an, da"s
das Feld $z(\x)$ durch die Variations-Hamiltonfunktion gau"sverteilt
ist. Dieses gilt nach D.S. Fisher \cite{dsf:86} zumindestens f"ur
$T=0$. Durch diese Annahme k"onnen wir das Wick-Theorem verwenden
(Anhang \ref{a:cos}).
\begin{equation*}
  \begin{split}
    0 &= \left<\left[ \int d^Dx -V_0 \cos \left(\frac{2\pi}{\as}
          z(\x)\right) - \frac 1 2 m^2z^2(\x) \right] \cdot \int d^Dy\
      m z^2(\vec y)\right>_{\H_0,C}\\
    &= - \int d^Dx\ d^Dy\ \left<\left[V_0 \cos \left(\frac{2\pi}{\as}
          z(\x)\right) + \frac 1 2 m^2z^2(\x) \right] \cdot m z^2(\vec
      y)\right>_{\H_0,C} \\
    &=
    - \int d^Dx\ d^Dy\ m \left<z(\x)z(\vec y)\right>^2_{\H_0} \left\{
        m^2 - V_0 \left(\frac{2\pi}{\as}\right)^2 e^{- \frac{2\pi^2}{a_\perp^2}
          \left<(z(\x))^2\right>_{\H_0} }\right\}
  \end{split}
\end{equation*}
Hinreichend f"ur die Erf"ullung dieser Gleichung ist das Verschwinden
des Ausdrucks in der geschweiften Klammer. Hierbei ist zu beachten,
da"s $\left<z^2(\x)\right>_{\H_0}$ eine Funktion von $m$ ist. Wir
wollen hier nat"urlich wieder ein unordnungsgemitteltes Ergebnis
haben. Daher m"ussen wir hier den unordnungsgemittelten Erwartungswert
von $z^2$ verwenden. Somit lautet die Selbstkonsistenzgleichung
\begin{equation}
  \label{var:skg}
  m^2 = V_0 \left(\frac{2\pi}{\as}\right)^2 e^{- \frac{2\pi^2}{\as^2}
    \overline{\left<(z(\x))^2\right>}_{\H_0} }.
\end{equation}

Wir berechnen $\overline{\left<z^2(x)\right>}_{\H_0}$ mit Hilfe der
replizierten Hamiltonfunktion.
\begin{equation*}
  \overline{\left<z^2(x)\right>}_{\H_0} = \lim_{n\rightarrow 0}
  \text{Tr }\left((z^\alpha)^2 e^{-\frac {\H^n_0}T}\right)
\end{equation*}
Hierzu entwickeln wir die replizierte Hamiltonfunktion bis zur zweiten
Ordnung in $z^\alpha$. Dieses liefert unter der Annahme, da"s das $z$
gau"sverteilt ist, ein korrektes Ergebnis. Des weiteren transformieren
wir noch die Hamiltonfunktion in die Fourierdarstellung, da diese
diagonal in den Impulsen $\q$ ist.
$$\H^n(\q) = \int \frac{d^Dq}{(2\pi)^D}\ \sum_{\alpha,\beta} \frac 12
\underbrace{\left((\Gamma q^2+ m^2)\delta^{\alpha,\beta} - \frac 2 T
    R''(0)n\right)}_{=:(G^{-1}(\q\,))^{\alpha\beta}}z_{\q}^\alpha
z_{-\q}^\beta$$ 
Die Mittelung besteht nun im wesentlichen in der Invertierung der
Matrix $G^{-1}$ (Anhang \ref{a:matrix}).
Wir erhalten
\begin{equation}
  \begin{aligned}[b]
    \label{var:z2}
    \overline{\left<z^2(x)\right>}_{\H_0} &= \int
    \frac{d^Dq}{(2\pi)^D}\ \lim_{n\rightarrow 0} T 
    \cdot G^{\alpha\alpha}(q)\\
    &= \int \frac{d^Dq}{(2\pi)^D}\ \left\{ \frac T{\Gamma q^2+m^2} - \frac
    {R''(0)}{\left(\Gamma q^2 + 
    m^2 \right)^2}\right\}
  \end{aligned}
\end{equation}
Der erste Summand ist dabei der thermische Beitrag zur Rauhigkeit. Aus
dem reinen Modell wissen wir, da"s der oberhalb von zwei
Oberfl"achendimensionen eine verschwindende Rauhigkeit liefert. Der zweite
Summand ist der Unordnungsanteil an der Rauhigkeit. Sein Beitrag ist
von der Ordnung $T^0$. Er liefert also auch bei $T=0$ einen
Beitrag.

So wie bis jetzt formuliert, wird die Unordnung ``nur''
st"orungstheoretisch behandelt. Dieses f"uhrt, wie wir bereits aus dem
Fall ohne periodisches Potential wissen, zu ungenauen Resultaten
aufgrund von vielen lokalen Minima der Hamiltonfunktion. Wir wollen daher
f"ur die Unordnung die Ergebnisse der Renormierungsgruppe
benutzen. Dabei wird die Kopplungskonstante durch eine effektive
Kopplungskonstante ersetzt. Diese effektive Kopplungskonstante ist
die Gr"o"se, die in der Renormierungsgruppe nach Ausintegration der
schnellen Freiheitsgrade definiert wurde. Sie ist dadurch
skalenabh"angig geworden, weshalb wir besser von einem Kopplungsterm
sprechen.

\section{Drei-Parameter-Approximation}
Eine besonders sch"one Methode, den effektiven Kopplungsterm zu
berechnen, besteht in der Drei-Parameter-Approximation von
T. Nattermann und H. Leschhorn \cite{hl:phd,tn+hl:91}. Hierbei wird
die funktionale Renormierung auf eine bestimmte Familie von Funktionen
eingeschr"ankt. Die so gefundenen Flu"sgleichungen f"ur die Parameter
dieser Funktionenfamilie sind elementar integrierbar, so da"s wir
nicht nur den Fixpunkt der Renormierungsgruppe erhalten, sondern auch
den Flu"s der Parameter dorthin. Ein Nachteil dieser Methode soll
nicht verschwiegen werden. Sie ist nur auf solche
Unordnungskorrelatoren anwendbar, deren Integral "uber die ganze
z-Achse nicht divergiert. Dieses ist f"ur das Random Bond Problem
gew"ahrleistet, wenn man den anf"anglichen Unordnungskorrelator durch
eine Gau"sfunktion darstellt.

Wir beginnen mit einer geeigneten Definition von $D(z)$
\begin{equation}
  \label{var:r_ansatz}
  D(z) = \frac \Delta \xi r\left(\frac z \xi \right) \qquad
  r(u) \approx \left\{ 
    \begin{array}{cl} 
      1 & |u| \ll 1 \\ 0 & |u| \gg 1
    \end {array} \right.
\end{equation}

Mit dieser Definition f"uhren wir zwei Parameter $\Delta$ und $\xi$
ein, die renormiert werden sollen. $\xi$ ist die Korrelationsl"ange
der Unordnung. $\Delta$ gibt die St"arke der Unordnung an. Die
Funktion $r$ ist kurzreichweitig. Wir verwenden hier sp"ater f"ur
Rechnungen eine Gau"sfunktion. Diesen Ansatz f"ur $D$ setzen wir nun
in die Flu"sgleichung (\ref{frg:fluss2}.b) f"ur $D$ aus der funktionalen
Renormierung ein. Wir erhalten
\begin{multline*}
  \frac 1 \xi\ r\left(\frac z\xi\right) \frac{\partial \Delta}{\partial
    l} - \left[ \frac \Delta{\xi^2}\ r\left(\frac z\xi\right) + \frac {
      \Delta z}{\xi^3}\ r'\left(\frac z\xi\right) \right]\frac {\partial
    \xi}{\partial l} = 
  [\epsilon - 4\zeta] \frac\Delta\xi\ r\left(\frac
    z\xi\right) \\ + 
  \frac {\zeta\Delta z}{\xi^2}\ r'\left(\frac z\xi\right)
  + \frac {\Delta^2}{\xi^6}\left[\left(r''\left(\frac
        z\xi\right)\right)^2 - 2\ r''\left(\frac z\xi\right)
    r''(0)\right] 
\end{multline*}

Von dieser Gleichung bestimmt man nun zwei verschiedene Momente
be\-z"ug\-lich $z$. Die einzelnen Summanden der Gleichung sind gerade
Funktionen von $z$, da der Unordnungskorrelator eine gerade Funktion
ist. Die Momente, die wir ausw"ahlen, m"ussen gerade sein, da die
ungeraden verschwinden. Aus diesen beiden Gleichungen f"ur die Momente
k"onnen wir nun die Flu"sgleichungen f"ur die Parameter $\Delta$ und
$\zeta$ gewinnen. Wir erhalten
\begin{equation}
  \label{var:fluss}
  \begin{aligned}[b]
    \frac {\partial \Delta}{\partial l} &= [\epsilon - 5\zeta +
    c_\Delta g] \Delta\\
    \frac {\partial \xi}{\partial l} &= [-\zeta + c_\xi g] \xi
  \end{aligned}
\end{equation}

Hier tauchen drei neue Gr"o"sen auf: $c_\Delta$, $c_\xi$ und $g$. Die
ersten beiden bestimmt man aus der folgenden Gleichung
\begin{equation}
  \label{var:const}
  c_\Delta + m c_\xi = \frac { \int du\ u^m r''(u)[r''(u)-2 r''(0)]}
  {\int du\ u^{m} r(u)}
\end{equation}
Dabei setzt man f"ur $m$ ganze gerade Zahl ein, die entsprechenden
Momente. $g$ ist eine dimensionslose Kopplungskonstante aus der
Kombination der beiden Parameter $\Delta$ und $\xi$: $g = \Delta /
\xi^5$. Aus den Gleichungen (\ref{var:fluss}) l"a"st sich eine
Flu"sgleichung f"ur $g$ herleiten:
\begin{equation}
  \frac {\partial g}{\partial l} = \epsilon g + \left[c_\Delta - 5
  c_\xi\right] g^2.
\end{equation}
Der Fixpunkt dieser Gleichung berechnet sich zu 
\begin{equation}
  g^* = \frac \epsilon {5c_\xi - c_\Delta}.
\end{equation}
Man kann die Flu"sgleichung nun von $l=0$ bis $l=\ln(\Lambda L)$
integrieren
\begin{equation*}
  \int_{g_o}^{\tilde g(l)} dg \frac 1 {\epsilon g\left(1-\frac
  g{g^*}\right)} = \int_0^l dl'  
\end{equation*}
und erh"alt
\begin{equation}
  \begin{aligned}[t]
  &\tilde g(l) = \frac {g_og^*e^{\epsilon l}}{g^* +g_o
  \left(e^{\epsilon l} - 1 \right)}\\
  &\text{ mit } g_o = g(l=0) = \frac {\Delta_0}{\xi_0^5}
\end{aligned}_.
\end{equation}
Damit k"onnen wir nun auch die Flu"sgleichungen f"ur $\tilde\Delta$ und
$\tilde \xi$ l"osen:
\begin{equation*}
  \begin{aligned}
    \int_{\tilde \Delta_0}^{\tilde \Delta(l)} \frac {d\Delta} \Delta &=
    \int_0^l dl'\ \left\{ \epsilon - 5\zeta + c_\Delta \frac
    {g_og^*e^{\epsilon l}}{g^* +g_o  \left(e^{\epsilon l} - 1
    \right)}\right\}\\  
    \int_{\tilde\xi_0}^{\tilde\xi(l)} \frac {d\xi} \xi &=
    \int_0^l dl'\ \left\{-\zeta + c_\zeta \frac {g_og^*e^{\epsilon
    l}}{g^* +g_o \left(e^{\epsilon l} - 1 \right)} \right\}
  \end{aligned}
\end{equation*}

Wir erhalten
\begin{equation}
  \label{var:effK}
  \begin{aligned}[b]
    \tilde \Delta \left(\Lambda L \right) &= \tilde\Delta_0  G^{\frac c
      {5-c}}(L)\\
    \tilde \xi \left(\Lambda L \right) &= \tilde\xi_0 G^{\frac 1 {5-c}}(L),
  \end{aligned}
\end{equation}
wobei wir zur Vereinfachung $c:=c_\Delta/c_\xi$ und
$$G(L) := 1 + \frac {g_o}{g^*}\left[\left(\Lambda L\right)^\epsilon - 1 
\right]$$
definiert haben.

Wir ersetzen nun die Kopplungskonstante $-R''(0)$ in (\ref{var:z2})
durch die aus dem Ansatz (\ref{var:r_ansatz}) herr"uhrende. Die
Gr"o"sen $\Delta$ und $\xi$ werden durch die unskalierten Versionen
der effektiven Gr"o"sen (\ref{var:effK}) ersetzt.  Wir k"onnen nun
$\overline{\left<z^2\right>}$ f"ur $T=0$ berechnen:
\begin{equation}
  \begin{aligned}[b]
    \overline{\left<z^2(x)\right>} &= \int \frac{d^Dq}{(2\pi)^D}\
    \frac {\Gamma^2\Lambda^\epsilon}{K_D} \frac
    {\Delta_0}{\xi_{0}^3} G^{\frac{c-3}{5-c}}\left(\frac 1
        q\right) \frac1{(\Gamma q^2 +
      m^2)^2}\\
    &=  \frac{\Delta_0}{\xi_{0}^3} \int^{\Lambda}_{1/L} dq\
    \frac {\Gamma^2 q^{D-1}}{(\Gamma q^2 +
    m^2)^2}G^{\frac{c-3}{5-c}}\left(\frac 1 
q\right)
  \end{aligned}
  \label{var:int1}
\end{equation}

$K_D$ ist hier wiederum das Volumen der $D$-dimensionalen
Einheitskugel im Impulsraum, dividiert durch $(2\pi)^D$. Die Masse $m$
bewirkt nun, da"s die kleinen Impulse abgeschnitten werden. Wir haben
also in einem unendlich gro"sen System einen unteren Cutoff von
$m/\sqrt\Gamma$. 
\begin{equation*}
  \begin{aligned}[b]
    \overline{\left<z^2(x)\right>} &\approx 
    \frac{\Delta_0}{\xi_{0}^3} \int^{\Lambda}_{m/\sqrt\Gamma} dq\
    q^{-1-\epsilon}G^{\frac{c-3}{5-c}}\left(\frac 1
      q\right)\\
    &= \frac {\xi_0^2} {2 c_\xi} \left[ G^{\frac 2{5-c}}\left(\frac
    {\sqrt \Gamma} m\right) -1 \right]
  \end{aligned}
\end{equation*}

Somit k"onnen wir nun die Selbstkonsistenzgleichung aufstellen. Es ist
im weiteren n"utzlich, die konkurrierenden physikalischen Beitr"age
durch miteinander vergleichbare L"angenskalen auszudr"ucken. 

Zuerst einmal haben wir bereits drei L"angen in der Gleichung
stehen, $\as$, $\xi_0$ und $1/\Lambda$ . $\as$ ist die Periode des
gittersimulierenden Potentials. Es ist somit eine L"ange in
$z$-Richtung. $\Lambda$ ist der obere Impulscutoff, somit ist
$1/\Lambda=a_{||}$ die Gitterkonstante in x-Richtung.

\begin{figure}[htb]
  \psfrag{z}{$z$}
  \psfrag{xi}{$\xi_0$}
  \centerline{\epsfig{figure=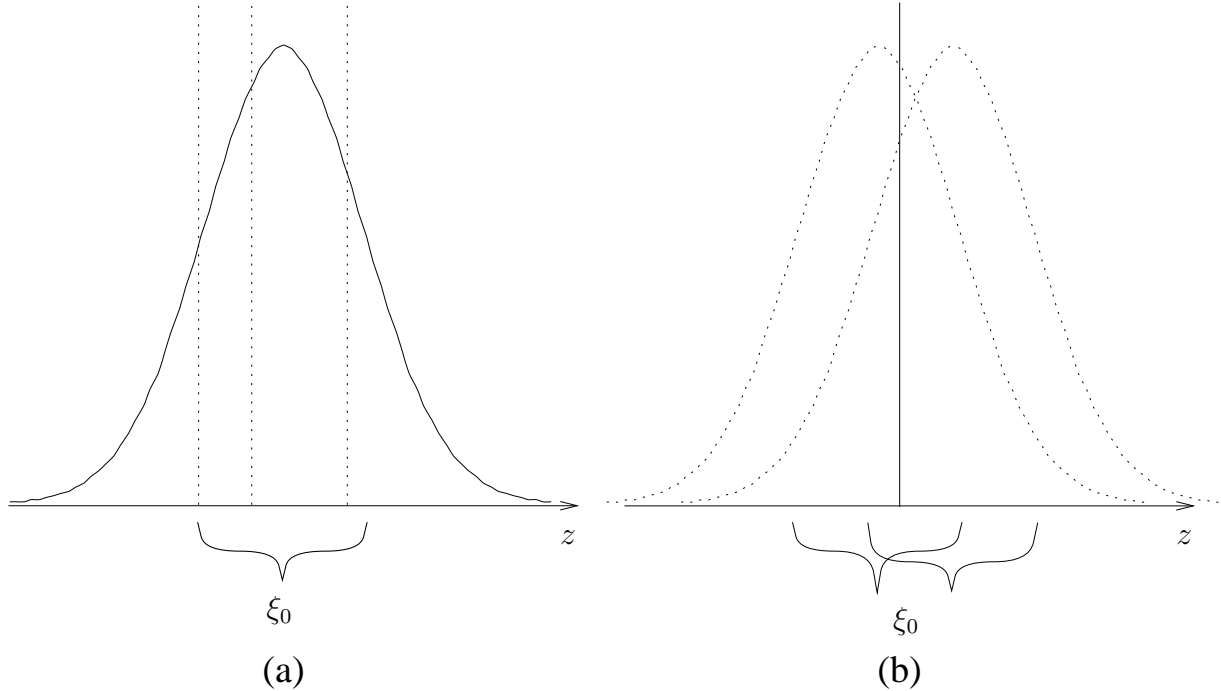,width=\textwidth}}
  \caption{\em Die Korrelationsl"ange $\xi_0$: Es gibt zwei
    Extremf"alle. Nehmen wir zun"achst (a) an, die Punktdefekte haben
    eine gewisse Breite $\xi_0$ (in der Abbildung durch die Gau"skurve
    dargestellt) und die Position der Wand sei sehr scharf
    definiert. In diesem Fall kann die Position der Wand im Intervall
    $\xi_0$ variieren ohne das sich der Energieanteil der Unordnung
    dramatisch "andert. Wir haben in der Skizze drei m"ogliche
    Positionen als gestrichelte Linien eingezeichnet. Im anderen Fall
    (b) sind Punktdefekte sehr schmal und die Wand hat eine Breite
    $\xi_0$. Auch hier ist es m"oglich, die Wand in diesem Intervall
    zu verschieben. In der Abbildung (b) ist die durchgezogene Linie
    die Position der Defektstelle. Die gestrichelten Linien zeigen
    exemplarisch zwei m"ogliche Positionen der Wand.}
  \label{var:wand-unordnugs-breite}
\end{figure}
$\xi_0$ ist das Maximum aus der Breite der Grenzfl"ache (bei einer
Blochwand ist das zum Beispiel die L"angenskala, auf der sich die Spins
um $180^o$ drehen) und der Korrelationsl"ange der Unordnung. Auch
dieses ist eine L"ange in $z$-Richtung. Diese L"ange gibt an, wie
stark man die Wand verschieben kann, ohne da"s sich der Energiebeitrag
der Unordnung drastisch "andert. Nehmen wir zum Beispiel an, wir haben
eine Wand der Breite $l$, die an einem Punktdefekt, welcher zun"achst
in der Mitte liegen soll, eine gewisse Energie $\Delta E$ gewinnt. Wir
k"onnen nun diese Wand um $\pm d/2$ verschieben. Der Punktdefekt
bleibt dadurch in der Wand, wodurch sich der Energiegewinn nicht
wesentlich "andert. Analog ist die Betrachtung einer {\em d"unnen}
Wand, welche mit Defektstellen endlicher Breite
wechselwirkt. (Abbildung \ref{var:wand-unordnugs-breite})

Die n"achste auftretende L"ange in unserem Problem beschreibt die
Unordnung. Es ist die sogenannte Larkin-L"ange $L_L$
\cite{larkin:jetp70}.  Das ist die L"angenskala, ab der die Unordnung
in der Art relevant wird, da"s man sie nicht mehr mit der
St"orungstheorie behandeln kann. Dieses ist dann der Fall, wenn die
Rauhigkeit $W$ der Wand von der Gr"o"senordnung der Gitterkonstante
$a_{||}$ wird. 
$$W^2(L) \simeq \frac {2 \Delta_o}{\xi_0^3 \epsilon}\left[\left(\frac L
    {a_{||}}\right)^\epsilon - 1\right]$$
Setzen wir nun $g_o$ und $g^*\propto\epsilon$ ein, so erhalten wir 
$$ a_{\perp}^2 \simeq \xi_0^2 \frac
{g_o}{g^*}\left(\frac{L}{a_{||}}\right)^\epsilon$$
Wir definieren daher die Larkin-L"ange als
\begin{equation}
  L_L:=a_{||} \left( \frac{g^*}{g_0}
  \left(\frac{a_{\perp}}{\xi_0}\right)^2 \right)^{1/\epsilon}  
\end{equation}

Wir sehen, da"s eine schwache Unordnung einer gro"sen Larkin-L"ange und
eine starke einer kleinen Larkin-L"ange entspricht. Weiterhin wird
die Larkinl"ange mit wachsender Korrelationsl"ange $\xi_0$
kleiner. Die Auswirkungen der Unordnung sind also um so st"arker, je
{\em breiter} die Defektstellen sind.

\begin{figure}[htbp]
    \centerline{\psfig{figure=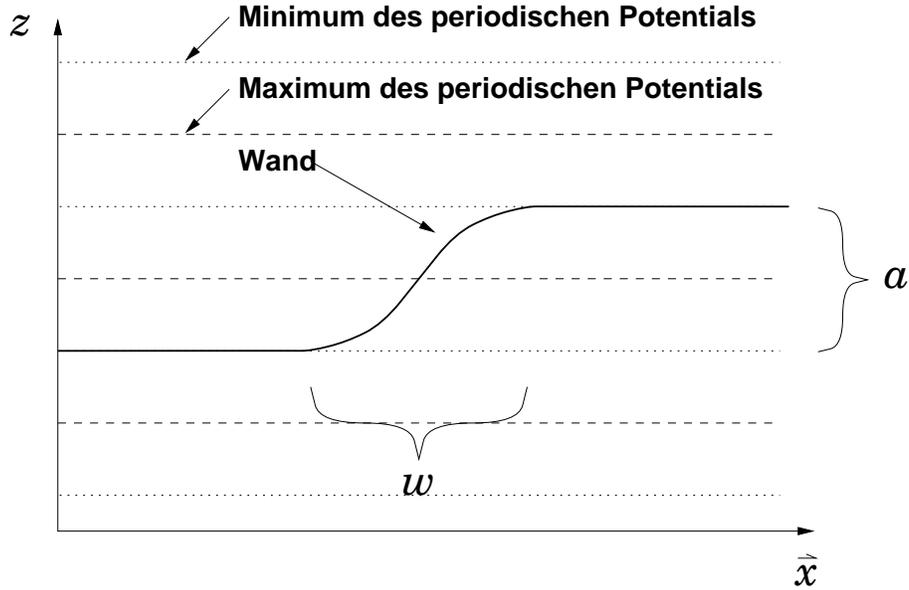,width=12cm}}
    \caption{\em Die L"angenverh"altnisse, die beim Wechsel der
      Oberfl"ache von einem Minimum zum n"achsten des Periodischen
      Potentials auftreten. Eine solche Stelle wird als Stufe oder Kink
      bezeichnet.}
\end{figure}
Die letzte L"angenskala ist dadurch gegeben, da"s die Wand eine
gewisse Distanz in $\x$-Richtung ben"otigt, um von einem Minimum des
periodischen Potentials zum n"achsten zu gelangen \cite{pn:91}. Diese
L"ange ist gepr"agt von der Konkurrenz zwischen der elastischen
Energie und dem periodischen Potential. Der elastische Term versucht
die Wand zu gl"atten, wodurch der "Ubergang zwischen zwei Minima des
periodischen Potentials in die L"ange gezogen wird. Das periodische
Potential wirkt diesem entgegen. Der Energieverlust durch dieses
Potential ist proportional zur Breite dieser Stufe. Die typische
Breite der Stufe ist nun dadurch bestimmt, da"s die Energiebeitr"age
der beiden Terme von der gleichen Gr"o"se sind. Der Beitrag der
elastischen Energie ist proportional zur Breite $w$, multipliziert mit
dem typischen Gradienten zum Quadrat $(\as/w)^2$. Mit den Vorfaktoren
$V_0$ und $\Gamma$ erhalten wir also
\begin{equation}
  w \sim \as \sqrt{\frac \Gamma {V_0}}
\end{equation}

Zuletzt skalieren wir noch den Variationsparameter $m$ zu einer
dimensionslosen Gr"o"se $p$ um. Wir dividieren dazu $m$ durch den
Vorfaktor der Exponentialfunktion auf der rechten Seite von Gleichung
(\ref{var:skg}) und definieren dieses als $p^2$.
\begin{equation}
  p := m \frac{\as}{2\pi} \frac 1 {\sqrt{V_0}}
\end{equation}
Der physikalisch relevante Bereich, in dem $p$ liegen kann, liegt
zwischen 0 und 1. Bei $p=1$ entspricht die Variations-Hamiltonfunktion
$\H_0$ einer Entwicklung des periodischen Potentials in der
urspr"unglichen Hamiltonfunktion $\H$ bis zur
quadratischen Ordnung in $z$. Werte von $p$, die gr"o"ser als 1
sind, w"urden bedeuten, da"s das periodische Potential die Rauhigkeit
unserer Wand st"arker unterdr"uckt als das Parabelpotential
der Variations-Hamiltonfunktion. Dieses ist physikalisch unsinnig: Der
Energieverlust pro Wandfl"ache f"ur eine rauhe Wand bleibt im
periodischen Potential immer endlich. Im Parabelpotential wird
er hingegen f"ur beliebig gro"se Auslenkungen beliebig gro"s.

Wir haben nun alle Gr"o"sen, um die Selbstkonsistenzgleichung
(\ref{var:skg}) aufzuschreiben:

\begin{eqnarray}
  \label{var:skg2}
  p^2 &= \exp\left( -\frac{\pi^2}{c_\xi}
  \left(\frac{\xi_0}{\as}\right)^2
  \left\{\left[1 + \left(\frac{\as}{\xi_0}\right)^2
  \left(\frac{L_L}{\ap}\right)^{-\epsilon} \left[\left(\frac w{2\pi\ap
        p}\right)^\epsilon - 1  \right]\right]^{\frac 2{5-c}} -1
  \right\}\right)_. \nonumber \\ \qquad
\end{eqnarray}

Die auftretenden Gr"o"sen sind nun entweder dimensionslos, wie $p$ und
die Konstanten $c$ und $c_\xi$, oder L"angenverh"altnisse. Die L"angen
werden in den nat"urlichen Einheiten des Systems gemessen. Dieses sind
die Gitterkonstanten in der entsprechenden Richtungen: $L_L$ und $w$
werden mit $\ap$, $\xi_0$ mit $\as$ gemessen.

Wir m"ussen jetzt noch die Konstanten $c_\xi$ und $c$ mit Hilfe von
Gleichung (\ref{var:const}) bestimmen. Um diese exakt zu bestimmen,
w"are es notwendig, die exakte Fixpunktfunktion f"ur $r(u)$
einzusetzen. Diese ist jedoch analytisch nicht bestimmbar und liegt
somit nur numerisch vor. Wir m"ussen also eine geeignete N"aherung
f"ur diese Funktion finden. Hieraus folgt, da"s die Konstanten nun von
der Wahl der Momente $m$ abh"angt. Wir setzen an:
\begin{equation*}
  \tilde r^*(u) = \exp\left(-\frac{u^2}2\right)
\end{equation*}
Der Nachteil dieser Funktion ist, da"s sie im Gegensatz zur
tats"achlichen Fixpunktfunktion am Ursprung analytisch ist. Die
richtige Fixpunktfunktion zeigt hier eine Divergenz in der vierten
Ableitung. Man sollte daher $m \neq 0$ w"ahlen.  Die Abh"angigkeit
von $m$ bei der gen"aherten Fixpunktfunktion ist allerdings sehr
gering. Mit den Momenten 2 und 4 erhalten wir
\begin{equation*}
  \begin{aligned}
    c &= 0.1885\dots\\
    c_\xi &= 2.1105\dots
  \end{aligned}
\end{equation*}
Der hieraus resultierende Rauhigkeitsexponent
\begin{equation*}
  \zeta = \frac 1 {5-c} \epsilon \approx 0.2078 \epsilon
\end{equation*}
kommt recht nahe an Fishers Ergebnis \cite{dsf:86} von
$0.2083\epsilon$.

\section{Auswertung der Selbstkonsistenzgleichung}
\psfrag{ll}{$L_L/\ap$}
\psfrag{xi}[r]{$\xi_0/\as$}
\psfrag{w}[r]{$w/\ap$}

Wir wollen nun die Selbstkonsistenzgleichung (\ref{var:skg2})
analysieren und deren L"osungen finden. Zun"achst stellen wir ein paar
allgemeine "Uberlegungen an. Wir betrachten die urspr"ungliche
Form der Selbstkonsistenzgleichung (\ref{var:skg}). Mit dem
reskalierten Variationsparameter $p$ lautet sie
\begin{equation*}
  p^2 = \exp\left(-\frac {2\pi^2}{\as^2}
    \overline{\left<z^2(\x)\right>} \right)_.
\end{equation*}
Weil ihr Argument immer negativ ist, ist die Exponentialfunktion
ihrerseits immer kleiner als eins. Somit ist die Bedingung, da"s f"ur
eine physikalische sinnvolle L"osung $p$ kleiner als eins sein mu"s,
immer erf"ullt. Bei der Berechnung von $\overline{\left<z(\x)\right>}$
wurde eine N"aherung gemacht, die nicht allgemein g"ultig ist. Im
Integral "uber die Impulse (\ref{var:int1}) ist die untere
Integralgrenze durch die Masse modifiziert worden. Dieses ist jedoch nur dann
sinnvoll, wenn $m/\sqrt\Gamma$ kleiner als die obere
Abschneidefrequenz der Impulse ist. F"ur gr"o"sere Werte wird die
N"aherung extrem schlecht und bewirkt zum Beispiel, da"s das Argument
der Exponentialfunktion in der Selbstkonsistenzgleichung
(\ref{var:skg2}) negativ werden kann, wodurch L"osungen $p > 1$
folgen. Die Bedingung $m/\sqrt\Gamma < \Lambda$  "ubersetzt sich in 
die Bedingung
\begin{equation}
  p \leq \frac w {2\pi\ap}
  \label{var:bed2}
\end{equation}
f"ur die neu definierten Gr"o"sen. Bei Gleichheit ist der Wert der
Parameter erreicht, bei dem die Exponentialfunktion in Gleichung
(\ref{var:skg2}) gleich eins wird.

\begin{figure}[htb]
  \centerline{\epsfig{figure=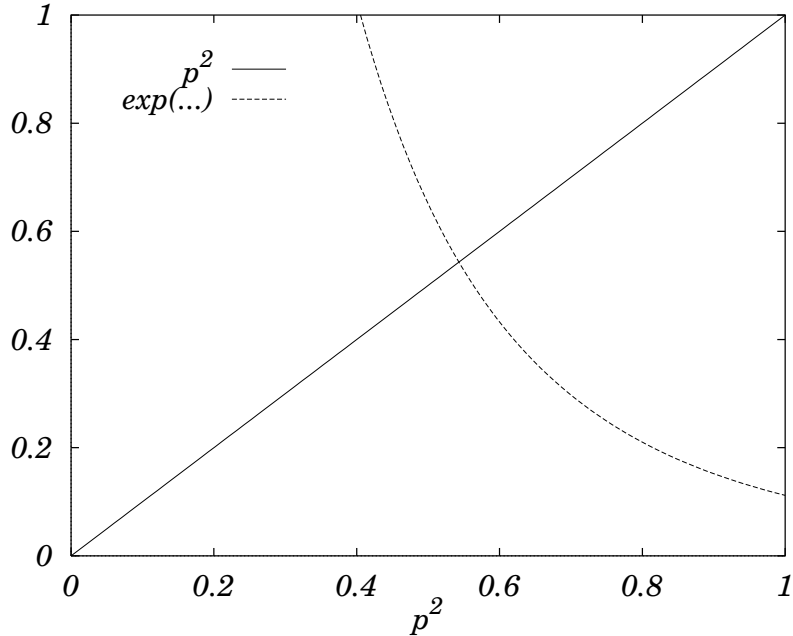,width=12cm}}
    \caption{\em Die Abbildung zeigt das typische Aussehen der
      Selbstkonsistenzgleichung f"ur Dimensionen $D>4$. Hier wurde
      speziell $D=5$, $w/\ap=4$, $L_L/\ap = 2$ und $\xi_0/\as=5$
      gew"ahlt. Auf der 
      Abszisse wird $p^2$, auf der Ordinate $p^2$ beziehungsweise die
      Exponentialfunktion aufgetragen. Der Schnittpunkt der beiden
      Kurven ist die L"osung der Gleichung.}
    \label{var:groesser4}
\end{figure}

Zun"achst wollen wir den Fall $D > 4$ betrachten. $\epsilon$ ist in
diesem Fall negativ, so da"s die Exponentialfunktion f"ur alle $p$
analytisch ist. Sie f"allt mit $p$ aufgrund des negativen Vorzeichens
im Exponenten und konvergiert gegen null f"ur $p \rightarrow
\infty$. Die Gleichung hat somit also immer eine L"osung $p>0$. Ein
Beispiel f"ur konkrete Werte der vorkommenden Gr"o"sen zeigt Abbildung
\ref{var:groesser4}. Die Grenzfl"ache ist f"ur diesen Fall also immer
glatt, wie wir erwartet haben.

Kommen wir nun zum Fall $D < 4$. In diesem Fall ist $\epsilon$
gr"o"ser als null. Das Argument der Exponentialfunktion ist an der
Stelle $p = 0$ nicht analytisch. Es divergiert gegen $-\infty$ f"ur $p
\rightarrow 0$, wodurch die Exponentialfunktion selber null wird. $p =
0$ ist also immer eine L"osung des Systems. Es verschwinden alle
Ableitungen der Funktion im Grenzwert $p \rightarrow 0$. Es kann also
nicht passieren, da"s eine weitere L"osung aus der Null
``herauswandert'', wie es zum Beispiel bei der Mean-Field Behandlung
des reinen Ising-Systems der Fall ist.

F"ur gr"o"serwerdende $p$ hat man das Problem, das irgendwann die
Bedingung (\ref{var:bed2}) nicht mehr erf"ullt ist. In diesem Fall ist
die N"aherung zu verwerfen und man mu"s das Integral numerisch behandeln.

\begin{figure}[htb]
  \centerline{\epsfig{figure=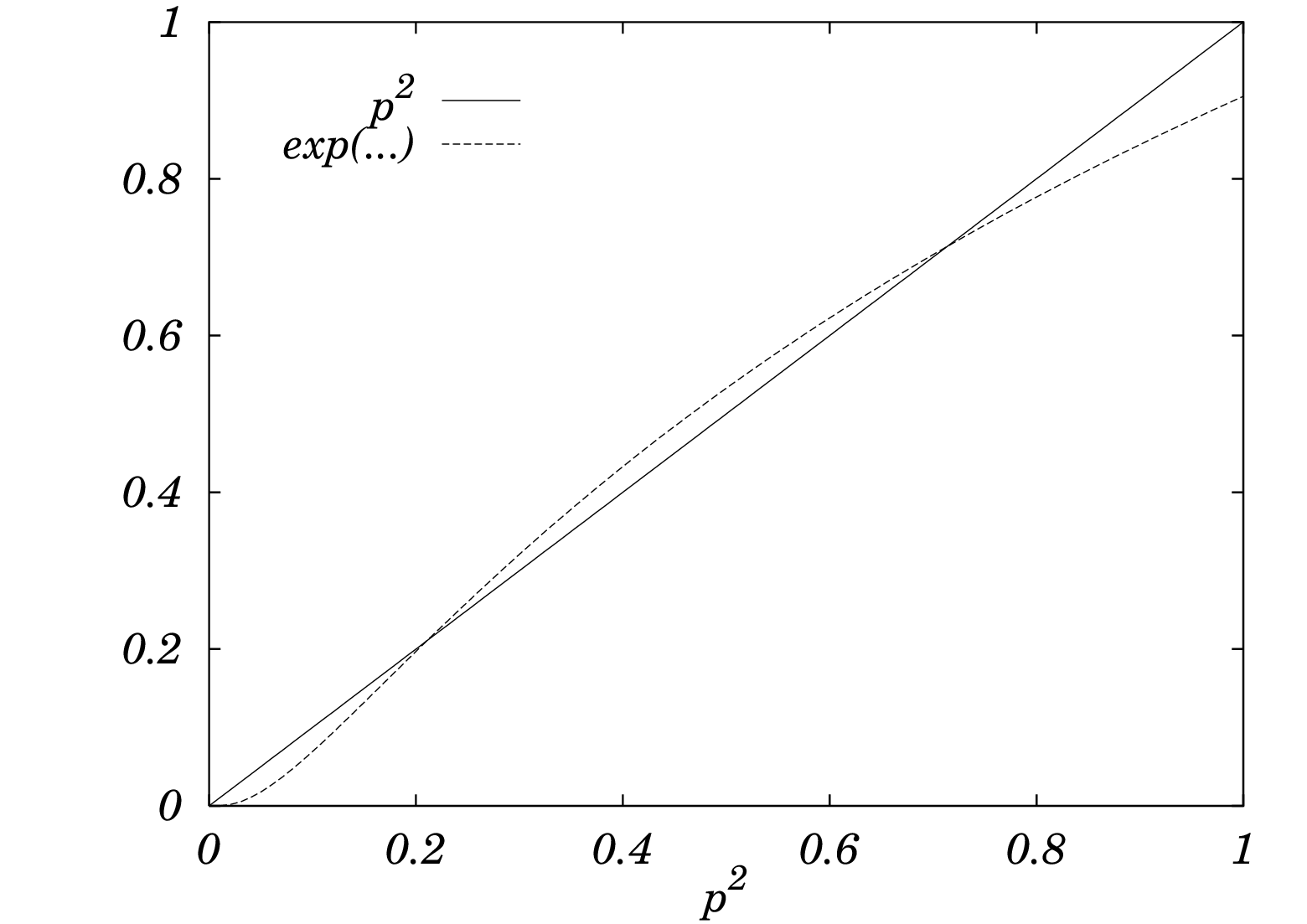,width=12cm}}
  \caption{\em Hier ist eine Situation f"ur $D < 4$ gezeigt. Die
    Auftragungsart ist dieselbe wie in Abbildung
    \ref{var:groesser4}. Hier hat sich das Verhalten der
    Exponentialfunktion drastisch ge"andert, so da"s es bis zu drei
    Schnittpunkten gibt.}
  \label{var:kleiner4}
\end{figure}

Das typische Aussehen der Selbstkonsistenzgleichung zeigt Abbildung
\ref{var:kleiner4}. Dort ist ein Fall gezeigt, wo die Gleichung drei
L"osungen hat, eine L"osung f"ur $p=0$ und zwei f"ur
nichtverschwindende $p$. Die $p = 0$ L"osung existiert f"ur alle
m"oglichen Werte der Parameter $L_L$, $\xi$ und $w$. Die Existenz der
anderen beiden Werte $p_1 < p_2$ ist von diesen Parametern
abh"angig. Die Wahl der Parameter bestimmt wie ``schnell'' bzw. steil
die Exponentialfunktion w"achst. 

Wir wollen nun untersuchen, wie sich die L"osungen $p_1$ und $p_2$ bei
kleinen Variation der Parameter $L_L$, $\xi_0$ und $w$ ver"andern.
Erh"ohen wir $w$, dieses entspricht einem schw"acheren periodischen
Potential, so wird $p_2$ kleiner und $p_1$ gr"o"ser.  Wir schlie"sen
daraus, da"s die gr"o"sere der beiden L"osungen, also $p_2$, die
physikalische L"osung ist: Ein kleineres $p$ unterdr"uckt die
Rauhigkeit weniger als ein gro"ses. Weiterhin bewirkt ein schw"acheres
periodisches Potential eine gr"o"sere Rauhigkeit als ein
starkes. Dieses beiden Tatsachen treffen f"ur $p_2$ zu.  Analog
ist das Verhalten bei "Anderung von $\xi_0$ und $L_L$. Erh"oht man
$\xi_0$ oder senkt man $L_L$, so wird in beiden F"allen $p_2$ kleiner
und $p_1$ gr"o"ser. Dieses pa"st ebenfalls in das physikalische Bild:
Eine st"arkere Unordnung sollte eine gr"o"sere Rauhigkeit und somit ein
kleineres $p$ zur Folge haben. Bei beiden F"allen erh"oht man die
Unordnung, einmal, indem man ihren ``Einflu"sbereich'' vergr"o"sert
($\xi_0$ erh"oht) und zum anderen indem man den numerischen Wert der
Unordnung erh"oht ($L_L$ kleiner macht).

Der Phasen"ubergang ist genau die Wertekombination von $\xi$, $L_L$
und $w$ bei der eine L"osung $p \neq 0$ zum ersten mal
auftritt. Betrachten wie die Gleichung (\ref{var:skg}), so stellen
wir fest, da"s, wenn die Exponentialfunktion f"ur irgendeinen Wert der
Masse $m$ ungleich null ist, wir mit einem gen"ugend gro"sen $V_0$
einen Schnittpunkt zwischen dieser und $p^2$ erreichen k"onnen. Die
Frage nach der Existenz eines Phasen"ubergangs reduziert sich
hierdurch auf die Frage, unter welchen Bedingungen die
Exponentialfunktion f"ur alle $p$ den Wert null annimmt.  Dieses ist
wiederum genau dann der Fall, wenn das Argument der
Exponentialfunktion nach $-\infty$ divergiert. Die Divergenz mu"s
wegen dem oben gesagten unabh"angig f"ur alle $p$ auftreten. Dieses ist
der Fall, wenn die Larkin-L"ange $L_L$ gegen null geht oder die
Korrelationsl"ange $\xi_0$ divergiert. Beides entspricht einer
unendlich gro"sen Unordnung und ist somit nicht realisierbar.  Eine
andere M"oglichkeit w"are eine Divergenz der L"ange $w$. Man kann
dieses durch $V_0 \rightarrow 0$ oder $\Gamma \rightarrow \infty$
erreichen. Es handelt sich also auch hier um nicht allzu sinnvolle
Grenzwerte, da zum einen kein periodisches Potential vorhanden ist und
zum anderen die Wand unendlich steif und somit immer glatt ist.

\begin{figure}[htb]
  \centerline{\epsfig{figure=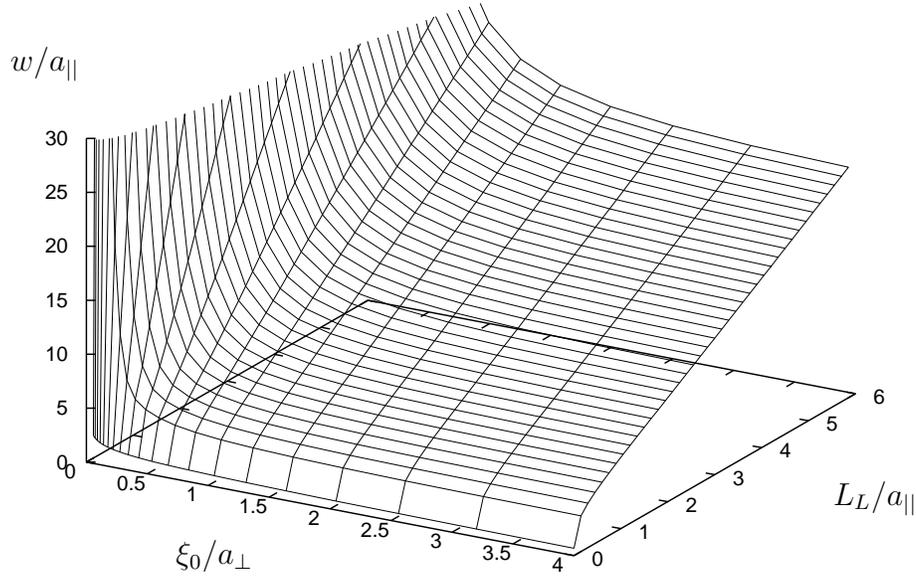,width=14cm}}
  \caption{\em Das Phasendiagramm: unterhalb der dargestellten Fl"ache
    ist der Ordnungparameter $p \neq 0$. Dieses bedeutet, da"s sich das
    System in der glatten Phase befindet. Oberhalb der eingezeichneten
    Fl"ache ist das System rauh. Bei der Berechnung dieses Diagramms
    wurde das Integral (\ref{var:int1}) numerisch ausgewertet.} 
  \label{var:phasen}
\end{figure}

Beschr"anken wir uns auf endliche, von null verschiedene Werte
der Parameter, so ist die Exponentialfunktion au"ser bei $p=0$ von
null verschieden. Somit gibt es f"ur diese Wahl der Parameter immer
einen durch das periodische Potential $V$ getriebenen Phasen"ubergang.

Wir stellen nun die L"ange $w$, die die St"arke des periodischen
Potentials beschreibt, als Funktion der anderen beiden L"angen $\xi_0$
und $L_L$ dar. F"ur fest vorgegebene Werte der Korrelations- und
Larkin-L"ange bestimmen wir den maximalen Wert $w(\xi_0,L_L)$ bei dem
es in Abbildung \ref{var:kleiner4} noch einen Schnittpunkt zwischen
den eingezeichneten Kurven gibt. In diesem Fall tangiert die
Exponentialfunktion die Gerade.
Dieses maximale $w$ stellt somit einen Phasen"ubergang zwischen einer
glatten Phase, bei der $p \neq 0$ ist, und einer rauhen mit $p=0$ dar. Der
Phasen"ubergang ist diskontinuierlich, das hei"st, $p$ springt von
einem endlichen Wert auf null. 

Hieraus ergibt sich ein Phasendiagramm, wie es Abbildung
\ref{var:phasen} zeigt. Die dargestellte Fl"ache ist die Grenzfl"ache
zwischen der rauhen (oberhalb) und der glatten Phase (unterhalb). Wir
sehen, da"s eine st"arkere Unordnung, das hei"st eine kleinere
Larkin-L"ange $L_L$, eine st"arkeres periodisches Potential ben"otigt
(kleineres $w$), um in eine glatte Phase zu gelangen. Bemerkenswert
ist die Abh"angigkeit des Phasen"ubergangs von der Korrelationsl"ange
$\xi_0$. Ist diese gro"s gegen"uber der Gitterkonstanten des
periodischen Potentials $\as$, so ist die Abh"angigkeit sehr
gering. F"ur Werte $\xi < \as$ zeigt sich demgegen"uber eine sehr
starke Abh"angigkeit. In diesem Fall reicht schon ein
verh"altnism"a"sig kleines Potential, um die Wand zu gl"atten. Ist die
Korrelationsl"angen von der Gr"o"senordnung des Gitterkonstante oder
gr"o"ser, so scheint die Unordnung einen Teil des periodischen
Potentials ``herauszumitteln'', weshalb der numerische Wert $V_0$
gr"o"ser sein mu"s, um eine glatte Wand zu erhalten.

Bei Random Field Unordnung haben wir die M"oglichkeit, das
Fixpunktergebniss der Unordnung zur Auswertung von Gleichung
\ref{var:skg} zu verwenden. Wir erhalten eine Flu"sgleichung, die
keinen Parameter enth"alt, welcher die St"arke der Unordnung genauer
charakterisiert. 
\begin{equation*}
  m^2=V_0 \left(\frac {2\pi}{\as}\right)^2\cdot \exp \left({-
  \left(\frac{2\pi}{\as}\right)^2 \frac {\epsilon}{4\zeta}
  \left(\left(\frac{\Lambda 
      \sqrt{\Gamma}}m \right)^{2\zeta}-1\right)}\right)
\end{equation*}
Die "Uberlegungen bei Random Bond Unordnung bez"uglich des
periodischen Potentials sind weiterhin g"ultig. Somit haben wir auch
hier einen durch $V_0$ getriebenen Phasen"ubergang. F"ur ein $V_0$,
welches kleiner als ein kritischen $V_c$ ist, ist die Oberfl"ache
rauh, f"ur ein $V_0$ gr"o"ser als $V_c$ glatt. Dieser Phasen"ubergang ist
ebenfalls diskontinuierlich.

Wir wollen nun versuchen, ob man auch aus der
Renormierungsgruppenrechnung einen Phasen"ubergang herleiten
kann. Wir "andern dazu die replizierte Hamiltonfunktion (\ref{repH})
derart ab, da"s wir noch einen Masseterm $\int d^Dx\ 1/2m^2z^2(\x)$
einf"ugen. 
\begin{multline}
  {\cal H}^n = \int d^Dx \left\{  \sum_\alpha \frac 1T \left[ \frac
      \Gamma {2} \left(\nabla z^\alpha(\vec x)\right)^2 
      + \frac 12 m^2 (z^{\alpha}(\x))^2
      - V_0\cos\left(\frac {2\pi}{\as} z^\alpha(\vec x) \right) \right]
  \right. \\ 
  \left. - \frac 1 {2T^2} \sum_{\alpha,\beta}R\left(z^\alpha(\vec x)
      -z^\beta(\vec x)\right)\right\} 
\end{multline}

Mit dieser neuen Hamiltonfunktion f"uhren wir nun die funktionale
Renormierung genau wie in Kapitel \ref{frg} durch. Das Einzige, was
sich "andert, ist der zu verwendende Propagator.
\begin{equation*}
  \frac 1 {\Gamma q^2} \rightarrow \frac 1 { \Gamma q^2 + m^2}
\end{equation*}
Wir ersetzen in Gleichung (\ref{frg:int2}) die beiden Propagatoren
und fahren mit denselben N"aherungen fort. Letztendlich bleibt ein
Integral der Art
\begin{equation*}
  \label{mas:int}
  \int d^Dq\ \frac 1 {(\Gamma q^2 + m^2)^2} 
\end{equation*}
"ubrig. Dieses Integral ist uns schon aus der Variationsrechnung
bekannt. Es bewirkt, da"s die untere Impulsabschneidefrequenz sich bei
hinreichend gro"ser Masse $m$ auf $m/\sqrt{\Gamma}$ erh"oht. Die
Flu"sgleichungen (\ref{frg:fluss}.b) und (\ref{frg:fluss}.c) "andern
sich hierdurch der Art, da"s die Terme, die die zweiten Ableitungen
von $V$ beziehungsweise $R$ enthalten, in diesem Fall wegfallen.
\begin{equation}
  \begin{split}
    \frac {\partial\tilde R}{\partial l} &= (\epsilon - 4\zeta)\tilde
    R + \zeta \tilde z\tilde R ' + \frac 1 2 \frac{K_D}{\Gamma^2}
    \Lambda^{-\epsilon}\left[\tilde R^{''2}(\tilde z)-2\tilde
    R''(\tilde z) \tilde R''(0)\right] \Theta\left(\frac \Lambda {e^l} -
    \frac m {\sqrt{\Gamma}} \right)\\ \frac {\partial\tilde
    V}{\partial l} &= (2 - 2\zeta) \tilde V + \zeta \tilde z \tilde V
    ' - \frac 1 2 \frac{K_D}{\Gamma^2} \Lambda^{-\epsilon}\tilde
    V''(\tilde z)\tilde R''(0)\Theta\left(\frac \Lambda {e^l} - \frac m
    {\sqrt{\Gamma}} \right)
  \end{split}
\end{equation}
$\Theta$ ist die Stufenfunktion, die f"ur positive Argumente den Wert 1
annimmt und sonst null ist. Sie bewirkt hier die Fallunterscheidung
zwischen massivem und masselosen Fall. F"ur $l \rightarrow \infty$
f"allt der letzte Term bei einer endlichen Masse immer weg.

Was bedeutet das nun f"ur das System? Ohne die Masse $m$ liefert
das Integral (\ref{frg:int2}) im thermodynamischen Limes f"ur den Grenzwert
$l \rightarrow \infty$ die divergenten Beitr"age, welche die Unordnung
als auch das peridische Potential auf nichttriviale Art
renormieren. Diese Divergenzen treten mit Masse nicht mehr auf, da die
Impulse nun nach unten beschr"ankt sind. Das bedeutet, da"s die
Renormierung bei einem endlichen Wert des Flu"sparameters $l$
stoppt. Dieses $l$ entspricht einer endlichen L"angenskala, ab der
sich die effektiven Kopplungskonstanten des Systems nicht mehr
"andern. Die Oberfl"ache bleibt in diesem Fall also glatt.

Die Frage, die noch zu kl"aren bleibt, ist, welchen Wert hat die Masse
$m$? Der Term, der uns diesen leifern kann, ist das periodische Potential. In
einer einfachen St"orungsrechnung w"urde uns die zweite Ableitung an
der Stelle $z=0$ diesen Term liefern
\begin{equation*}
  m^2= \left.\frac {\partial^2}{\partial z^2} V(z)\right|_{z=0} = 
  V_0\left(\frac {2\pi}{\as}\right)^2_.
\end{equation*}

Wir verwenden hier, um die Masse zu bestimmen, ein mit dem massiven
Propagator renormiertes $V_0$. Das Vorgehen ist dasselbe wie in
Abschnitt  \ref{frg:fp} mit dem Unterschied, da"s die Masse
mitgenommen wird. Somit enden wir bei der Flu"sgleichung f"ur $V_0$
\begin{equation}
   \frac {\text d V_0}{\text d q_0} = \frac 1 2 \Lambda^{2\zeta}
   \left(\frac{2\pi}{\as}\right)^2\epsilon q_0^{-2\zeta-1}
   V_0\ \Theta\left(q_0 - \frac m {\sqrt{\Gamma}} \right)
\end{equation}
In der Stufenfunktion wurde $e^l=\Lambda/q_0$ ersetzt. Die folgende
Impulsintegration wird wieder durch die Masse nach unten
begrenzt. Somit erhalten wir
\begin{equation}
  m^2=V_0 \left(\frac {2\pi}{\as}\right)^2\cdot \exp \left({-
  \left(\frac{2\pi}{\as}\right)^2 \frac {\epsilon}{4\zeta}
  \left(\left(\frac{\Lambda 
      \sqrt{\Gamma}}m \right)^{2\zeta}-1\right)}\right)
\end{equation}

Mit diese Gleichung ist nun die Masse $m$ selbstkonsistent zu
bestimmen. Es ist die Geleichung, welche wir erhalten w"urden, wenn
wir im vorangehenden Kapitel den unordnungsgemittelten Erwartungswert
$\overline{\left<z^2\right>}$ nicht mit der drei Parameter
Approximation, sondern `nur' mit dem Fixpunktresultat der
Renormierungsgruppe f"ur die Unordnung, berechnet h"atten. Was diese
Methode nicht liefert, ist der Crossover von kleinen zu gro"sen
L"angenskalen. Diesen erh"alt man nur dann, wenn man die
Flu"sgleichungen der Funktionalen Renormierung simultan
integriert. Das ist aber nicht ohne weitere Vereinfachungen, wie eben
der drei Parameter Approximation, m"oglich. Somit hat sich der Kreis
geschlossen.

\chapter{Zusammenfassung}

In dieser Arbeit wurde eine elastische Oberfl"ache in einem
ungeordneten Medium mit einem zus"atzlichen periodischen Potential
untersucht. 
Ziel dieser Arbeit war es, die Existenz eines
Phasen"ubergangs zwischen einer rauhen und einer glatten Phase zu
beleuchten. Die obere kritische Dimension f"ur dieses System ist D=4,
genau wie im Fall ohne periodisches Potential \cite{dsf:86}. Oberhalb
dieser Dimension ist sowohl das periodische Potential als auch die
Unordnung irrelevant.

Zun"achst wurde das Problem mithilfe der funktionalen Renormierung
untersucht. Dabei wird die Abweichung von der kritischen Dimension als
kleiner Parameter $\epsilon = 4 - D$ genommen, und nach diesem bis zur
f"uhrenden Ordnung entwickelt. Hieraus resultieren Flu"sgleichungen
f"ur die Temperatur $T$, das periodische Potential und die
Unordnung. Das System wird durch einen $T=0$ Fixpunkt beschrieben.
Die Flu"sgleichungen f"ur die anderen beiden Gr"o"sen sind partielle
Differentialgleichungen.  Die Beschreibung der Unordnung wird nicht
durch das periodische Potential beeinflu"st. Somit ergibt sich
hierf"ur derselbe Fixpunkt wie bei D.S. Fisher \cite{dsf:86}. Die
Flu"sgleichung des periodischen Potentials l"a"st sich auf einer
Flu"sgleichung der Kopplungskonstante $V_0$ dieses Potentials
reduzieren. Diese wiederum liefert, eine stabilen Fixpunkt
$V_0=0$. Somit ist das periodische Potential immer irrelevant.

Eine zweite Methode, mit der das Problem behandelt wurde, ist die
Variationsmethode. Das periodische Potential wurde durch ein
quadratisches ersetzt. Die dazugeh"orige Kopplungskonstante $m$ wurde
dabei so bestimmt, da"s ein geeignetes Variations-Funktional minimal
wird. Die Unordnung wurde bei dieser Rechnung weiterhin mithilfe der
Renormierungsgruppe behandelt. Diese wurde derart abge"andert, da"s
der Funktionenraum, in dem sich der Korrelator der Unordnung befindet,
derart beschr"ankt wurde, da"s sich die Renormierungsgruppe auf die
Renormierung dreier Konstanten reduziert. Die drei Konstanten sind
die Temperatur $T$, die st"arke der Unordnung $\Delta$ und die Breite
der Defektstellen $\xi$. Diese Approximation bietet sowohl Vor- als
auch Nachteile: Der wichtigste Vorteil ist, da"s nun der "Ubergang
(Crossover) von kleinen zu gro"sen L"angenskalen berechnet werden
kann. Der gr"o"ste Nachteil ist, da"s hiermit nur lokale
Unordnungskorrelatoren, wie sie bei der Random Bond Unordnung
auftreten, betrachtet werden k"onnen. Das Ergebnis dieser Rechnungen
ist, da"s es Unterhalb von $D=4$ Oberfl"achendimensionen einen
Phasen"ubergang erster Art zwischen einer rauhen und einer glatten
Phase gibt. Was diese Rechnung nicht liefert, ist eine untere
kritische Dimension, ab der das periodische Potential immer irrelevant
ist. 

In einem weiteren Abschnitt wurde versucht, auch aus der
Renormierungsgruppenrechnung einen Phasen"ubergang herzuleiten. Zu
diesem Zweck ist die Hamiltonfunktion durch einen Masseterm erweitert
worden. Dieser Term bewirkt, da"s die Renormierung auf einer durch die
Masse bestimmten L"an\-gen\-ska\-la zum erliegen kommt. Versucht man nun
diese Masse aus den im urspr"unglichen Modell vorhanden Gr"o"sen zu
extrahieren, so endet man wiederum bei einer selbstkonsistent zu
l"osenden Gleichung. Im Gegensatz zur Variationsrechnung enth"alt
diese nur die Fixpunktwerte der Unordnung und keinen Crossover.

\chapter*{Anhang}
\addcontentsline{toc}{chapter}{Anhang}
\setcounter{chapter}{1}
\setcounter{equation}{0}
\markboth{Anhang}{}

\section{Das Variations-Funktional}
\label{a:funktional}
Angenommen, wir haben zwei Hamiltonfunktionen $\H$ und $\H_0$. Wir
ben"otigen ein Funktional dieser beiden Funktionen, welches f"ur $\H
= \H_0$ sein Minimum annimmt.

Durch die beiden Hamiltonfunktionen werden zwei Verteilungsfunktionen
definiert: 
\begin{xalignat}{2}
  \rho &= \frac {e^{-\H/T}} Z &\quad Z &= \text{Tr}\ e^{-\H/T} \nonumber\\
  \rho_0 &= \frac {e^{-\H_0/T}} {Z_0} &\quad Z_0 &= \text{Tr}\
  e^{-\H_0/T} \nonumber
\end{xalignat}
Wir betrachten nun 
\begin{equation*}
  \rho_0 \cdot \left( \ln {\rho_0} - \ln \rho \right) = \rho_0 \cdot
  \ln \frac {\rho_0}\rho_.
\end{equation*}
Weiterhin benutzen wir die bekannte Ungleichung f"ur Logarithmen
\begin{equation*}
  \ln x \leq x-1 \quad \Rightarrow \quad \ln x \geq 1 - \frac
  1 x
\end{equation*}
und erhalten
\begin{equation}
  \label{var:1}
  \rho_0 \cdot \left( \ln {\rho_0} - \ln \rho \right) \geq \rho_0 -
  \rho.
\end{equation}
Die Spur "uber die rechte Seite dieser Ungleichung verschwindet, da
beides auf eins normierte Verteilungsfunktionen sind. Somit
ist die Spur "uber die linke Seite immer gr"o"ser gleich null.

\begin{equation*}
  \begin{aligned}
    0 &\leq \text{Tr}\left\{\rho_0 \cdot \left( \ln {\rho_0} - \ln \rho
      \right) \right\}\\
    &= \text{Tr}\left\{\rho_0 \cdot \left( -\ln Z_0 - \frac {\H_0}T +
        \ln Z + \frac {\H}T \right) \right\}\\
    &= \ln Z - \ln Z_0 + \frac 1 T \left<\H - \H_0\right>_{\H_0}
  \end{aligned}
\end{equation*}
$\left<\dots\right>_{\H_0}$ bedeutet, da"s der Erwartungswert mit $\H_0$
gebildet wird. Multiplizieren wir dieses Funktional mit $T$ und nehmen
nur die Anteile mit, die von $\H_0$ abh"angen, so erhalten wir ein
f"ur die Variationsrechnung geeignetes Funktional
\begin{equation*}
  \boxed{
    F^{\text{var}} = F_0 + \left<\H - \H_0\right>_{\H_0}},
\end{equation*}
wobei $F_0 = -T \ln Z_0$ die Freie Energie des durch $\H_0$
beschriebenen Systems ist. Die Ungleichung (\ref{var:1}) ist genau
dann eine Gleichung, wenn $\rho = \rho_0$ ist. Somit ist das
Variations-Funktional  genau dann minimal, wenn $\H = \H_0$
gilt. Ebenfalls sind in diesem Fall die drei Freien Energien
$F^{\text{var}}$, $F_0$ und $F$ gleich.

\section{Erwartungswert von $\cos(2\pi z/a)$}
\label{a:cos}
Wir wollen den Erwartungswert von $\cos(2\pi z/a)$ bez"uglich einer in
$z$ quadratischen Hamiltonfunktion $\H$ berechnen. 
\begin{equation*}
  \left<\cos\left(\frac {2\pi}a z\right)\right>_{\H} = \frac
  {\text{Tr}\ \cos(2\pi z/a) e^{-\H/T}} {\text{Tr}\ e^{-\H/T}} 
\end{equation*}
Der dazugeh"orige Boltzmannfaktor hat eine gau"ssche Form. Somit ist
das Wick Theorem anwendbar. Zu Beginn zerlegen wir den Kosinus in
seine Reihenentwicklung
\begin{equation*}
  \left<\cos\left(\frac {2\pi}a z\right)\right>_{\H} = \sum_n \frac 1
  {2n!} (-1)^n \left(\frac {2\pi}a\right)^{2n} \left<z^{2n}\right>_{\H} 
\end{equation*}
Der Erwartungswert von $\left<z^{2n}\right>_{\H}$ berechnet sich nun
mit dem Wick Theorem: Man mu"s die Erwartungswerte aller m"oglichen
Paarungen von $z$ berechnen. Da alle $z$ gleich sind ist dieses
besonders einfach, n"amlich gleich $\left<z^2\right>^n$ multipliziert
mit der Anzahl der M"oglichkeiten, die $z$ zu paaren. F"ur das erste
$z$ hat man $2n-1$ m"ogliche Partner. Es bleiben $2n-2$ $z$
"ubrig. Somit gibt es f"ur das zweite Paar $2n-3$
M"oglichkeiten. F"uhrt man dieses Verfahren f"ur alle $z$ durch, so
erh"alt man insgesammt $(2n-1)!!$ M"oglichkeiten.
\begin{equation*}
   \left<\cos\left(\frac {2\pi}a z\right)\right>_{\H} = \sum_n \frac
   1{2n!} (-1)^n \left(\frac {2\pi}a\right)^{2n}
   (2n-1)!!\left<z^2\right>^n_{\H} 
\end{equation*}
Zuletzt wollen wir versuchen, durch geschickte Umformungen die Summe
wieder aufzusummieren. Wir zerlegen die Fakult"at in zwei Doppelfakult"aten.
Die ungerade der beiden k"urzen wir gegen die bereits
vorhandene Doppelfakult"at. Die gerade zerlegen wir in $2^nn!$.
\begin{equation*}
  \left<\cos\left(\frac {2\pi}a z\right)\right>_{\H} = \sum_n \frac
  1{n!} \left(-\frac 12 \left(\frac {2\pi}a\right)^2
    \left<z^2\right>_{\H} \right)^n 
\end{equation*}
Diese Summe identifizieren wir als Exponentialfunktion.
\begin{equation*}
  \boxed{
    \left<\cos\left(\frac {2\pi}a z\right)\right>_{\H} =
    \exp\left(-\frac 12 \left(\frac{2\pi}a\right)^2
      \left<z^2\right>_{\H} \right)}
\end{equation*}

Analog berechnet sich der Erwartungswert
\begin{equation*}
  \left<z_1^2\cos\left(\frac {2\pi}a z_2\right)\right>_{\H,C}.
\end{equation*}
Weil hier eine Kumulate berechnet wird, ist zu beachten, da"s
zun"achst die beiben $z_1$ mit jeweils einem $z_2$ gepaart werden. Man
erh"alt:
\begin{equation*}
  \boxed{ 
    \left<z_1^2\cos\left(\frac {2\pi}a z_2\right)\right>_{\H,C} =
    \left<z_1z_2\right>^2_{\H,C}\left(\frac {2\pi}a\right)^2
    \exp\left(-\frac 12 \left(\frac{2\pi}a\right)^2
    \left<z_2^2\right>_{\H,C} \right)}
\end{equation*}

\section{Matrixinversion}
\label{a:matrix}

Gegeben sei eine quadratische $n\times n$ Matrix 
\begin{equation}
  \label{inv.1}
  M^{\alpha\beta}=a \delta^{\alpha\beta}+b,
\end{equation}
von der die inverse Matrix zu bestimmen ist. Wir nehmen an, da"s die
Inverse dieselbe Struktur besitzt.
\begin{equation}
  \label{inv.2}
  (M^{-1})^{\alpha\beta} = x \delta^{\alpha\beta} + y
\end{equation}

Es mu"s gelten
\begin{equation*}
  I_n = M \cdot M^{-1} = M^{-1} \cdot M,
\end{equation*}
wobei $I_n$ die $n\times n$ Einheitsmatrix ist.
F"ur die einzelnen Elemente der Matrix gilt daher
\begin{equation*}
  \delta^{\alpha\beta} = (M\cdot M^{-1})^{\alpha\beta} = \sum_\gamma
  M^{\alpha\gamma}(M^{-1})^{\gamma\beta}.
\end{equation*}
Nach Einsetzten von (\ref{inv.1}) und (\ref{inv.2}) erhalten wir
\begin{equation*}
  \begin{aligned}[b]
    \delta^{\alpha\beta} &= \sum_\gamma \left\{ a\cdot x\cdot
      \delta^{\alpha\gamma} \cdot\delta^{\gamma\beta} + a \cdot y
      \cdot\delta^{\alpha\gamma} + b \cdot x \cdot\delta^{\gamma\beta} + b
      \cdot y \right\}\\
    &= a\cdot x\cdot \delta^{\alpha\beta} + a\cdot y + b \cdot x + n \cdot
      b\cdot y.
    \end{aligned}
\end{equation*}
Dieses Gleichungssystem hat die eindeutige L"osung
\begin{equation*}
  \begin{aligned}[b]
    x &= \frac 1 a\\
    y &= - \frac b { a\cdot (a + n\cdot b)}_.
  \end{aligned}
\end{equation*}

Somit lautet die gefundene Inverse 
\begin{equation*}
  \boxed{
  (M^{-1})^{\alpha\beta} = \frac 1a \delta^{\alpha\beta} - \frac
  b{a\cdot(a+ n\cdot b)}}_.
\end{equation*}

\end{document}